\pdfoutput=1
\documentclass[11pt,twoside,a4paper,cmspaper,final,collab]{cms-tdr}
\def\svnVersion{e8fa916}\def\svnDate{2023/06/27}\def\cmsCernNoTag{CERN-EP-2022-114}\def\cmsCernDate{\today}\def\cmsMessage{Published in the Journal of High Energy Physics as \href{http://dx.doi.org/10.1007/JHEP06(2023)130}{\doi{10.1007/JHEP06(2023)130}.}}
\begin{document}\cmsNoteHeader{HIG-20-004}

\newcommand{\ttH}{\ensuremath{\PQt\PAQt\PH}\xspace}
\newcommand{\bbH}{\ensuremath{\PQb\PAQb\PH}\xspace}
\newcommand{\ggH}{\ensuremath{\Pg\Pg\PH}\xspace}
\newcommand{\HH}{\ensuremath{\PH\PH}\xspace}

\providecommand{\cmsTable}[1]{\resizebox{\textwidth}{!}{#1}}
\newlength\cmsTabSkip\setlength{\cmsTabSkip}{1ex}

\cmsNoteHeader{HIG-20-004} 
\title{Search for nonresonant Higgs boson pair production in the four leptons plus two \texorpdfstring{\PQb}{b} jets final state in proton-proton collisions at \texorpdfstring{$\sqrt{s}=13\TeV$}{sqrt(s)=13 TeV}}
\date{\today}

\abstract{The first search for nonresonant production of Higgs boson pairs (\HH) with one \PH decaying into four leptons and the other into a pair of \PQb quarks is presented, using proton-proton collisions recorded at a center-of-mass energy of $\sqrt{s}=13\TeV$ by the CMS experiment. The analyzed data correspond to an integrated luminosity of 138\fbinv. A 95\% confidence level upper limit of 32.4 is set on the signal strength modifier $\mu$, defined as the ratio of the observed \HH production rate in the $\HH\to\PZ\PZ^{*}\PQb\PAQb\to 4\ell\PQb\PAQb$ decay channel to the standard model (SM) expectation. Possible modifications of the \PH trilinear coupling $\lambda_{\PH\PH\PH}$ with respect to the SM value are investigated. The coupling modifier $\kappa_{\lambda}$, defined as $\lambda_{\PH\PH\PH}$ divided by its SM prediction, is constrained to be within the observed (expected) range $-8.8\,(-9.8) < \kappa_{\lambda} < 13.4\,(15.0)$ at 95\% confidence level.}

\hypersetup{
pdfauthor={CMS Collaboration},
pdftitle={Search for nonresonant Higgs boson pair production in the four leptons plus two b jets final state in proton-proton collisions at sqrt(s) = 13 TeV},
pdfsubject={CMS},
pdfkeywords={CMS, Higgs}}

\maketitle

\section{Introduction}

The discovery of the Higgs boson (\PH) by the ATLAS and CMS Collaborations~\cite{Aad:2012tfa,Chatrchyan:2012ufa} is the first experimental proof of the predicted mechanism of electroweak symmetry breaking.
Further measurements of the \PH properties performed by both Collaborations were found to be compatible with the standard model (SM) predictions~\cite{Aad:2015zhl,Sirunyan:2017exp,CMS:2018lkl}.
A determination of the \PH self-coupling provides an independent test of the SM by probing the Higgs scalar field potential~\cite{Baglio:2012np}; this parameter can be obtained by measuring the production of \PH pair (\HH) at the LHC.

By exploiting different decay channels, studies of \HH production allow different regions of the kinematic parameters describing possible extensions of the SM and of the nonresonant invariant mass spectrum to be probed. A combination of different decay channels is therefore needed to achieve the best possible sensitivity to \HH production.

A wide variety of \HH decay channels and their combination were studied by the ATLAS~\cite{Aad:2014yja, Aad:2015uka, Aad:2015xja, Aaboud:2016xco} and CMS~\cite{Khachatryan:2016sey, Khachatryan:2015year, Sirunyan:2017isc,  Sirunyan:2017tqo, Sirunyan:2017djm, Sirunyan:2017guj, CMS:2020gsy} Collaborations: the most recent published results on the \HH combination~\cite{ATLAScomb, CMScomb}, with a data set corresponding to an integrated luminosity of about 36 \fbinv, set an observed (expected) upper limit of, respectively, 7 (10) and 22 (13) at the 95\% confidence level (CL) on the SM production cross section times the theoretical prediction.

\begin{figure}[htb!]
  \centering
  \includegraphics[width=0.80\columnwidth]{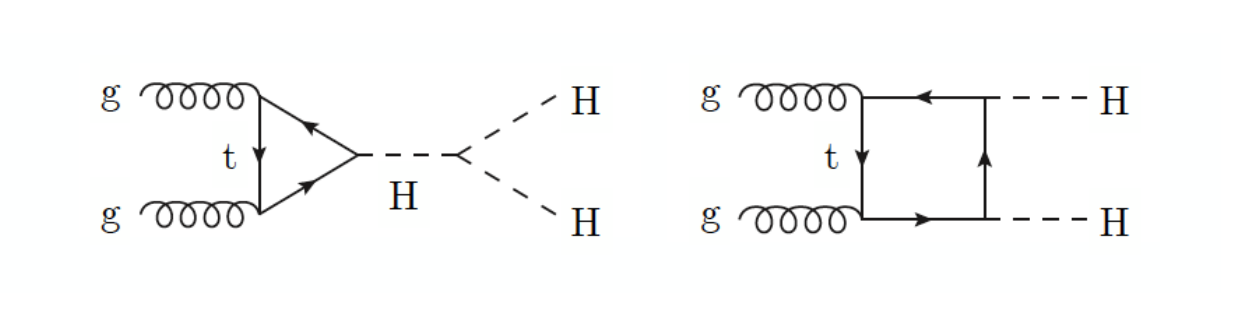}
   \caption{Feynman diagrams contributing to Higgs boson pair production via gluon-gluon fusion in the SM at LO.}
  \label{fig:ggF}
\end{figure}

The $\PH$ to four-lepton ($\PH\to \PZ\PZ^{*} \to 4\ell$, $\ell = \Pe, \mu$) decay channel is the rarest observed so far at the LHC but it has the largest signal-to-background ratio. 
The analysis presented in this paper focuses on gluon-gluon fusion \HH production where one \PH decays to a \PZ boson pair, which in turn decays into $4\ell$, and the other to a pair of $\PQb$ quarks ($\PQb\PAQb$), hadronizing into jets. The final state forms a clear signature granted by the presence of the four leptons, while the high branching fraction of the $\PQb\PAQb$ decay channel partially compensates for the small branching fraction of the $4\ell$ channel. The leading-order (LO) $\Pg\Pg \to \PH\PH$ Feynman diagrams are shown in Fig. \ref{fig:ggF}.

{\tolerance=800 The \HH SM production cross section via gluon-gluon fusion has been computed at next-to-next-to-LO (NNLO) in quantum chromodynamics (QCD), including next-to-next-to-leading logarithmic (NNLL) corrections and finite top quark mass ($m_{\text{top}}$) effects at next-to-LO (NLO). Its value is $\sigma_{\HH} =31.05^{+2.2}_{-5.0}\% (\text{scale uncertainties}) \pm 2.1\% (\text{PDF}) \pm 2.1\% (\alpha_S) ^{+4.0}_{-18.0}\% (m_{\text{top}}) \unit{fb}$ in proton-proton ($\Pp\Pp$) collisions at 13\TeV for a \PH mass of 125\GeV \cite{Baglio:2020wgt}, where PDF stands for parton distribution function. After accounting for the branching fraction ${\mathcal{B}(\PH\PH\to \PZ\PZ^{*} \PQb\PAQb \to 4\ell \PQb\PAQb)}$, the cross section is reduced to approximately 4.5\unit{ab} in the SM.\par}
Beyond the SM physics can significantly modify the cross section and the kinematic properties of \HH production, such as the $\PH\PH$ invariant mass distribution. In order to provide constraints on these effects, an electroweak chiral Lagrangian model~\cite{Heinrich:2017kxx,Buchalla:2018yce}, an effective field theory based approach that extends the SM with dimension-six operators, is used in an equivalent way as done in Ref.~\cite{Goertz:2014qta} but at NLO in QCD. This approach results in five couplings relevant for HH production, out of which the \PH trilinear coupling $\lambda_{\PH\PH\PH}$ is selected to perform scans of its allowed variations based on the results of this analysis. The variations of this coupling with respect to its SM prediction are parametrized by the coupling modifier, $\kappa_\lambda = \lambda_{\PH\PH\PH} / \lambda_\text{SM}$.

\section{The CMS detector}

The CMS detector features a superconducting solenoid of 6\unit{m} internal diameter, providing a magnetic field of 3.8\unit{T}. Within the solenoid volume are a silicon pixel and strip tracker, a lead tungstate crystal electromagnetic calorimeter, and a brass and scintillator hadron calorimeter, each composed of a barrel and two endcap sections. Forward calorimeters extend the pseudorapidity ($\eta$) coverage provided by the barrel and endcap detectors. Muons are detected in gas-ionization chambers embedded in the steel flux-return yoke outside the solenoid. Events of interest are selected using a two-tiered trigger system.
A detailed description of the CMS detector, together with a definition of the coordinate system used and the relevant kinematic variables, can be found in Ref.~\cite{Chatrchyan:2008zzk}.

\section{Data sets and simulations} 
\label{sec:datasets}

Data from $\Pp\Pp$ collisions collected by the CMS experiment in 2016--2018, at a center-of-mass energy of 13 TeV, are used for the analysis. Data-taking conditions vary year by year, and the samples used correspond to a total integrated luminosity of 138 \fbinv.
The triggers used for this analysis require the presence of a pair of loosely isolated leptons (muons or electrons), whose exact requirements depend on the data-taking year. 
Triggers requiring a triplet of low transverse momentum (\pt) leptons, and isolated single-electron and single-muon triggers are used as well to recover efficiency.
The overall trigger efficiency for events that satisfy the 4$\ell$ selection described below is greater than 98\%.
The trigger scheme used is the same as that used in the $\PH\to\PZ\PZ^{*}\to4\ell$ analysis previously published by the CMS Collaboration~\cite{h4lFullRun2}.

Signal samples of the SM $\Pg\Pg \to \PH\PH\to \PZ\PZ^{*} \PQb\PAQb \to 4\ell \PQb\PAQb$ process are generated at NLO in QCD using the \POWHEG v2~\cite{Alioli:2008gx,Nason:2004rx,Frixione:2007vw} generator with the implementation described in~\cite{Heinrich:2017kxx,Buchalla:2018yce}. 
Using the same model, beyond SM signal samples with alternative values of $\kappa_\lambda$ in steps of size 0.5 between $-10$ and $+10$ are generated. 
Single \PH production processes constitute the main SM background for this analysis. 
Descriptions of these processes are obtained using the \POWHEG v2 generator for the six main production modes: 
gluon fusion ($\Pg\Pg\PH$), vector boson fusion (VBF)~\cite{Nason:2009ai}, and associated production ($\PW\PH$, $\PZ\PH$, \bbH, and  \ttH~\cite{Hartanto:2015uka}). 
In the case of $\Pg\Pg\PH$, $\PW\PH$, and $\PZ\PH$, the \textsc{MiNLO} extension of \POWHEG is used~\cite{Campbell:2012am,Luisoni:2013kna} to increase the accuracy in the dijet phase space. 
The description of the decay of the $\PH$ to $4\ell$ is obtained using the \textsc{JHUgen} generator~\cite{Gao:2010qx} version 7.0.9. 
In the case of $\PW\PH$, $\PZ\PH$, \bbH, and \ttH, the $\PH$ is allowed to decay to $\PH \to \PZ \PZ^{*} \to 2\ell2$X where X could be any other particle from the Z decay so that $4\ell$ events where two leptons originate from the decay of associated vector bosons or top quarks are also taken into account. 
The hadronic shower propagation of parton-level events, referred to as showering, is then performed by allowing QCD emissions at all energies in the shower and by vetoing them afterwards according to the \POWHEG internal scale \cite{Nason:2004rx}. 

Several backgrounds include genuine nonresonant $\PZ\PZ^{*}$ events. Production of $\PZ\PZ^{*}$ via quark-antiquark annihilation is generated at NLO using \MGvATNLO with up to one extra parton, with appropriate settings to merge jet multiplicities~\cite{Frederix:2012ps}. As this simulation covers a large range of $\PZ\PZ^{*}$ invariant masses $m(\PZ\PZ^{*})$, dynamical factorization and renormalization scales have been chosen.
The simulated sample is rescaled by the ratio (denoted \textit{K}-factor) of NNLO to NLO cross sections, and additional NLO electroweak corrections~\cite{Grazzini:2015hta} are applied to the Monte Carlo (MC) sample as a function of $m(\PZ\PZ^{*})$.

The $\Pg\Pg \! \to \! \PZ\PZ^{*}$ process is simulated at LO with \MCFM~\cite{MCFM,Campbell:2013una} version 7.0.1. In order to match the $\Pg\Pg \! \to \mathrm{H} \to \! \PZ\PZ^{*}$ \pt spectra predicted at NLO, the showering for \MCFM samples is performed allowing only emissions up to the parton-level scale.
An exact calculation beyond the NLO does not exist for the $\Pg\Pg\to\PZ\PZ^{*}$ background, but it has been shown~\cite{Bonvini:1304.3053} that the soft collinear approximation can describe the background cross section and the interference term at NNLO. The NNLO \textit{K}-factor for the gluon fusion \PH production is obtained as a function of $m(\PZ\PZ^{*})$ using the HNNLO v2 MC generator~\cite{Catani:2007vq,Grazzini:2008tf,Grazzini:2013mca}.
Triboson production with at least one \PZ boson with leptonic decays is generated at NLO using \MGvATNLO, using similar settings.

The contribution from $\PZ$+jets events is estimated from data-driven techniques as described in Section~\ref{sec:redBkg}. Other small backgrounds, such as $\PQt\PAQt\PZ$, are estimated from simulation.

The \PYTHIA 8.2.30~\cite{Sjostrand:2015} package is used for parton showering, hadronization, and the underlying event simulation, with parameters set according to the CUETP8M1 tune~\cite{Khachatryan:2015pea} for the 2016 data-taking period and the CP5 tune~\cite{CP5} for the 2017-2018 data-taking periods. The NNPDF set of PDFs~\cite{NNPDF2015} is used, NNPDF3.0 (3.1) for the 2016 (2017-2018) periods.

The detector response is simulated using a detailed description of the CMS detector implemented using the \GEANTfour package~\cite{GEANT, Geant2}. Simulated events are reconstructed by the same algorithms used for the data. 
The simulated samples include additional interactions in the same and neighboring bunch crossings, referred to as pileup.
Simulated events are weighted to match the pileup distribution observed in the data; this procedure is performed separately for each year.

\section{Event reconstruction and selection}
\label{sec:eventselection}

The final state consists of at least two pairs of opposite-charged 
isolated leptons and at least two jets. The $4\ell$ selection is similar to that used in the CMS $\PH\to\PZ\PZ^{*}\to4\ell$ measurement described in Refs.~\cite{Sirunyan:2017exp,h4lFullRun2}. 

Events are reconstructed using a particle-flow algorithm~\cite{CMS-PRF-14-001} that aims at reconstructing and identifying each individual particle with an optimized combination of all subdetector information. 

The primary interaction vertex, from which the \PH candidates arise, is chosen to be the one corresponding to the hardest scattering in the event, evaluated using tracking information alone, as described in Ref.~\cite{CMS-TDR-15-02}.

Electrons are identified with a multivariate classifier, which includes observables sensitive to Bremsstrahlung along the electron trajectory, the geometrical and kinematical compatibility between the electron track and the associated energy cluster in the electromagnetic calorimeter, the shape of the electromagnetic shower, and variables that discriminate against electrons originating from photon conversions~\cite{Khachatryan:2015hwa}. 

In order to further suppress electrons from photon conversions and muons originating from in-flight decays of hadrons, the three-dimensional impact parameter significance is required to be smaller than four.

Muons are reconstructed by combining information from the silicon tracker and the muon system~\cite{Sirunyan:2018fpa}.
The muons are selected from the reconstructed muon track candidates by applying minimal requirements on the track in both the muon system and silicon tracker, and taking into account their compatibility with small energy deposits in the calorimeters.

Muons are required to be isolated from other particles in the event. The relative isolation is defined as
\begin{equation}
R_\text{iso} = \bigg[ \sum_{\substack{\text{charged} \\ \text{hadrons}}} \!\! \pt \, + \,
\max\big(0, \sum_{\substack{\text{neutral} \\ \text{hadrons}}} \!\! \pt
\, + \, \sum_{\text{photons}} \!\! \pt \, - \, \Delta\beta
\big)\bigg] \bigg/ \pt^{\ell},
\label{eq:iso}
\end{equation}
where the sums run over the charged and neutral hadrons, and photons, in a cone defined by $\Delta R \equiv \sqrt{\smash[b]{\left(\Delta\eta\right)^2 + \left(\Delta\phi\right)^2}} = 0.3$ around the lepton trajectory, where $\phi$ is the azimuthal angle in radians, and $\pt^{\ell}$ is the muon transverse momentum.
To minimize the contribution of particles originating from pileup interactions to the isolation calculation, charged hadrons are included only if they originate from the primary vertex. 
The $\Delta\beta$ term is a correction factor accounting for the energy deposits originating from pileup interactions~\cite{CMS:2020ebo}.
Muons with $R_\text{iso} < 0.35$ are considered isolated. A similar requirement is not applied to electrons, because the isolation information is included in the multivariate classifier.

The efficiency of the lepton reconstruction and selection is measured in both data and simulated samples in bins of lepton transverse momentum, $\pt^{\ell}$, and pseudorapidity, $\eta^{\ell}$, by using the tag-and-probe technique applied to leptonic Z boson decays~\cite{tagNprobe}. 
Correction factors, denoted \textit{scale factors} (SFs), are used to rescale the simulation so that it matches the data~\cite{h4lFullRun2}.
Leptons energy scale and resolution corrections are derived using the \JPsi meson and \PZ boson leptonic decays in bins of $\pt^{\ell}$ and $\eta^{\ell}$.

The lepton final-state radiation is recovered following an algorithm described in Ref.~\cite{Sirunyan:2017exp}.

Jets are reconstructed from particle-flow candidates using the anti-\kt clustering algorithm~\cite{Cacciari:2008gp}, as implemented in the \textsc{FASTJET} package~\cite{Cacciari:2011ma}, with a distance parameter of 0.4. To ensure a good reconstruction efficiency and to reduce the instrumental background as well as the contamination from pileup, loose identification criteria based on the multiplicities and energy fractions carried by charged and neutral hadrons are imposed on jets~\cite{CMS-PAS-JME-16-003}. Only jets in the tracker fiducial region ($\abs{\eta} < 2.4$) are considered.

Jet energy corrections are extracted from data and simulated events to account for the effects of pileup, non-uniformity of the detector response, and residual differences between the jet energy scale in the data and in the simulation.
The jet energy scale calibration~\cite{JINST6, Khachatryan:2016kdb, JESCalib} relies on corrections parametrized in terms of the raw (uncorrected) \pt and $\eta$ of the jet. In order to ensure that jets are well measured and to reduce the pileup contamination, all jets must have a corrected $\pt > 20\GeV$. 

Jets originating from the fragmentation of one \PQb quark are identified by using the \emph{DeepCSV} algorithm~\cite{BTAG2018}, which combines impact parameter significance, secondary vertex, and jet kinematical information to provide a final output discriminator whose score is computed using a deep artificial neural network. 
Data-to-simulation $\PQb$-tagging correction factors are applied to simulated events as a function of jet \pt, $\eta$, and flavour. 

A signal event must contain at least two \PZ boson candidates, each formed from pairs of isolated electrons or muons of opposite electrical charges; these leptons are selected according to the studies described in Ref.~\cite{Sirunyan:2017exp}.
Only reconstructed electrons (muons) with a $\pt > 7(5) \GeV$ are considered.
Among the $4 \ell$, the highest-\pt one must have $\pt > 20\GeV$, while the second-highest \pt one must have $\pt > 12\,(10)\GeV$ if it is an electron (muon).
All leptons are required to be separated by $\Delta R \left(\ell_1, \ell_2 \right) > 0.02$ from each other, and electrons are required to be separated from muons by $\Delta R \left(\Pe, \PGm \right) > 0.05$.

Within each event, all permutations of leptons giving a valid pair of \PZ boson candidates are considered. 
For each $\PZ\PZ$ candidate, the lepton pair with the invariant mass closest to the nominal $\PZ$ boson mass is denoted $\PZ_1$ and required to have a mass greater than 40\GeV. 
The other dilepton candidate is denoted $\PZ_2$ and must satisfy $12 < m_{\PZ_2} < 120$\GeV where the lower cut is used to discard low mass dilepton resonances while the upper threshold corresponds to the higher limit of the $\PZ_1$ mass range, as described in Ref.~\cite{Sirunyan:2017exp}.
All pairs of oppositely charged leptons that can be built from the $\PZ\PZ$ candidate, regardless of flavour, are required to satisfy a requirement on their invariant mass, $m_{\ell \ell'}$, to be larger than 4\GeV to suppress the QCD background with low transverse momentum leptons from hadron decays. 
If more than one $\PZ\PZ$ candidate passes the selection, the one with the highest value of the scalar \pt-sum of leptons is retained.

In order to search for \HH production, in addition to the $\PZ\PZ$ selections, events are required to have at least two jets, which are required to be separated from the leptons of the $\PZ\PZ$ candidate by $\Delta R(\text{lepton}, \text{jet})>0.3$ and to have $\pt > 20 \GeV$. At this stage, no other requirement on the value of the invariant mass or on the score of the  \emph{DeepCSV} $\PQb$-tagging algorithm for the jets is applied. 
The \ensuremath{\PQb\PAQb} candidate is formed using the two jets with the highest \PQb-tagging score.

\begin{figure}[htb!]
  \centering
  \includegraphics[width=0.48\columnwidth]{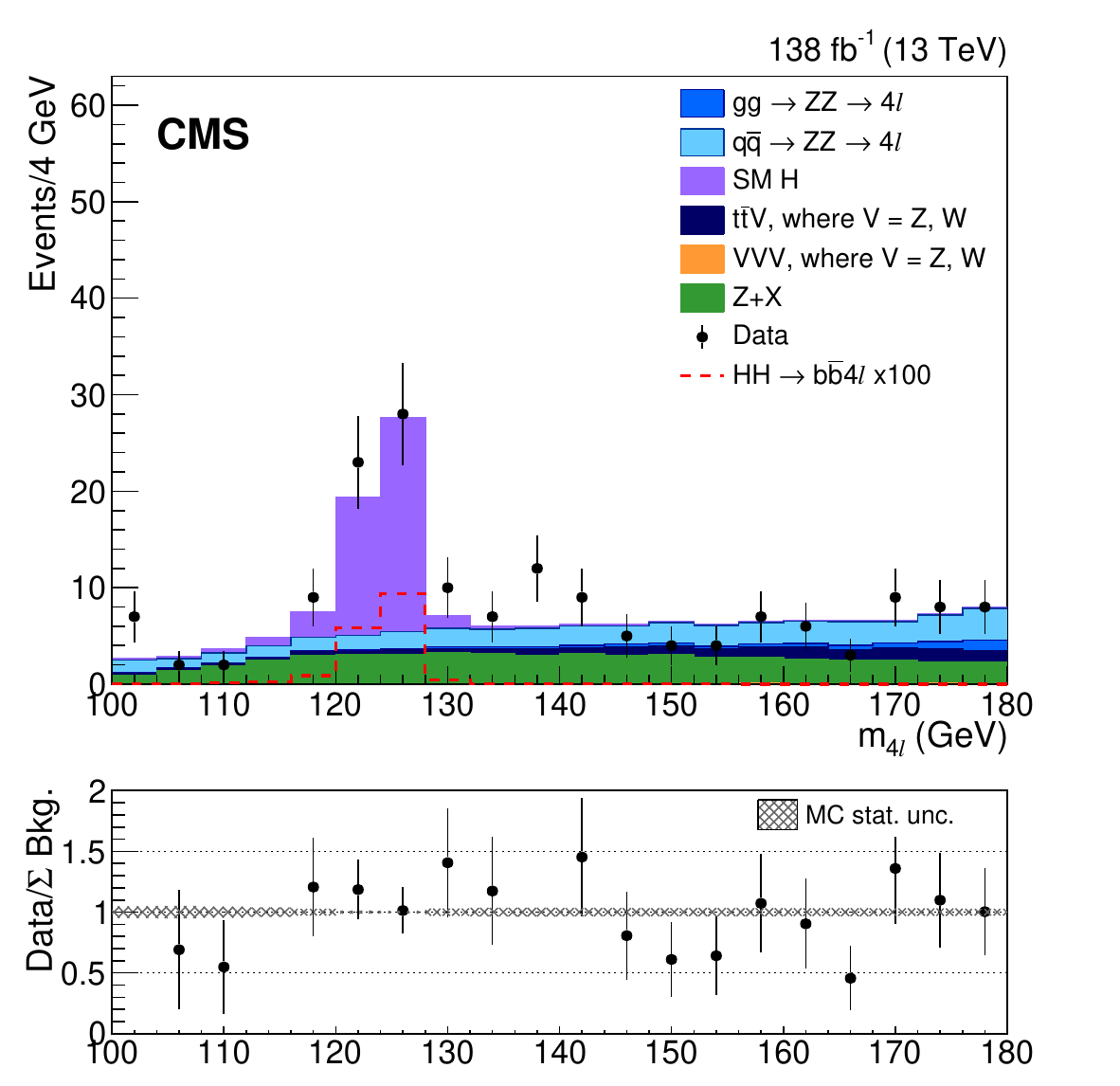}
  \includegraphics[width=0.48\columnwidth]{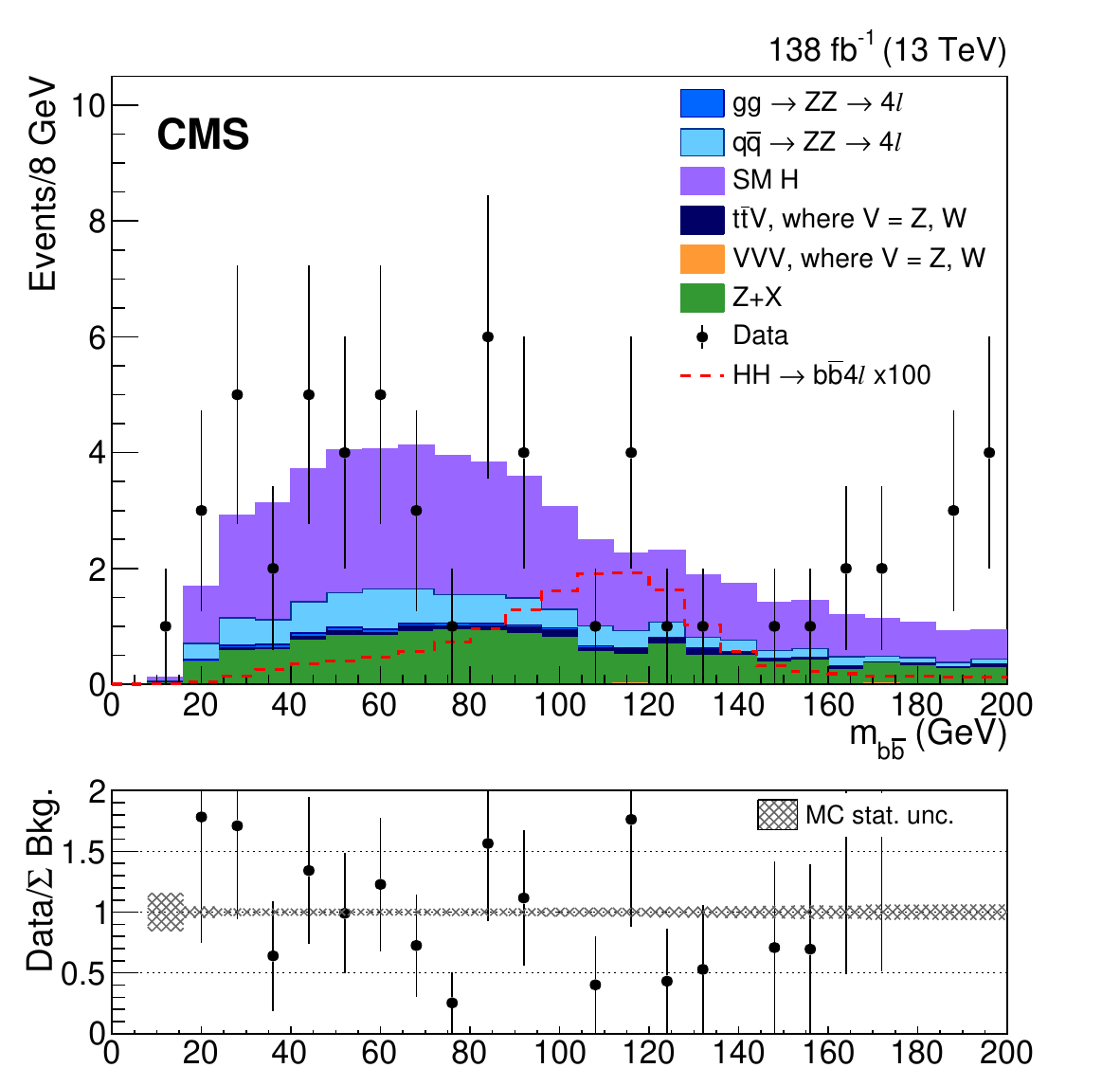}
  \caption{The $m(4\ell)$ (left) and $m(\PQb\PAQb)$ (right) distributions. The $m(\PQb\PAQb)$ plot is drawn after the selection: $115 < m(4\ell) < 135\GeV$. In the ratio plot, the grey dashed shades represent the relative statistical uncertainty on the total estimated background.}
  \label{fig:invariantmasses}
\end{figure}

The 4$\ell$ \emph{signal region} (SR) is defined by applying the additional requirement on the four-lepton invariant mass to be ${115<m(4\ell)<135\GeV}$. 
Events that do not satisfy this requirement are included in the 4$\ell$ \emph{control region} (CR).
The expected yields in the SR after the above selection are reported in Table~\ref{tab:yields}. 
Figure~\ref{fig:invariantmasses} shows the $m(4\ell)$ and $m(\PQb\PAQb)$ distributions in signal, estimated background components, and data.

\begin{table}[htbp]
  \centering
  \topcaption{Expected yields in the 4$\ell$ SR after additionally requiring at least two jets in the event. ``Others'' represents contributions from $\PQt\PAQt\PW$, $\PW\PW\PZ$, $\PW\PZ\PZ$, and $\PZ\PZ\PZ$ processes.}
  \cmsTable{
      \begin{tabular}{l c c c c c c c c c c c c}\hline
      \multicolumn{1}{c}{\textbf{Final state}} & \textbf{Signal} & \textbf{$\PQt\PAQt\PZ$} & \textbf{$\PQt\PAQt\PH$} & \textbf{\bbH} & \textbf{$\PZ\PZ$} & \textbf{$\Pg\Pg\PH$+VBF} & \textbf{$\PZ\PH$+$\PW\PH$} & \textbf{Others} & \textbf{$\PZ$+X} & \textbf{Total expected} & \textbf{Observed}\\
      \hline
      4$\mu$   & 0.056 & 0.58 & 0.68 & 0.16 & 3.75 & 12.35 & 1.61 & 0.04 & 3.87 & 23.10 & 29\\
      4$\Pe$       & 0.030 & 0.37 & 0.39 & 0.07 & 1.42 & 6.16  & 0.82 & 0.02 & 2.65 & 11.93 & 12\\
      2$\Pe$2$\mu$ & 0.082 & 0.95 & 0.91 & 0.20 & 4.93 & 15.56 & 2.08 & 0.10 & 7.22 & 32.03 & 33\\
      \hline 
      \end{tabular}
  }
\label{tab:yields}
\end{table}

\section{Background estimation}
\label{sec:bkgestimation}

\subsection{Irreducible background}

The irreducible background to the \HH signal is estimated from simulation. One of the main background contributions is the production of single \PH decaying into $\PZ\PZ^{*}$, whose description is provided using the \POWHEG v2 generator. For the gluon fusion \PH production mode, NNLO parton shower weights are applied. Another large contribution is the $\PQq\PAQq\to \PZ\PZ^{*}$ process, which is simulated at NLO and then reweighted with NNLO/NLO \textit{K}-factors to account for the missing higher orders in the cross section used. The $\Pg\Pg\to \PZ\PZ^{*}$ contribution too is estimated from MC by reweighting it with NNLO \textit{K}-factors. Other backgrounds, such as $\PQt\PAQt\PZ$ and $\PQt\PAQt\PW$, are also estimated from simulation.

\subsection{Reducible background}
\label{sec:redBkg}

The reducible background (Z+X) originates from processes that contain one or more non-prompt leptons (leptons not originating from the hard-interaction primary vertex). The main sources of non-prompt leptons are non-isolated electrons and muons coming from decays of heavy-flavour hadrons, misreconstructed jets (usually originating mostly from light-flavour quarks), and electrons from photon conversions. 

The number of expected events for these background processes is estimated by measuring the probabilities $f_{\Pe}$ and $f_{\PGm}$ for misidentified electrons and muons that do pass looser selection criteria to also pass the final selection criteria in selected samples of $\PZ(\ell\ell) + \Pe + 2$ jets and $\PZ(\ell\ell) + \PGm + 2$ jets events, where the additional lepton (\Pe or \PGm) satisfies a loose selection. 
The misidentification probabilities are evaluated using the requirement $\abs{m_{\text{inv}}(\ell_{1}\ell_{2}) - M_{\PZ}} < 7 \GeV$, to reduce the contribution from asymmetric conversions of photons, which is dominant at low dilepton masses. The misidentification rates are measured in bins of the \pt of the loosely-identified lepton and separately for the  barrel and the endcap regions. 

To estimate the Z+X yield in the SR, we define two control samples as subsets of $\PZ+2$ jets events that pass the Z candidate selection, 
requiring an additional pair of loose leptons of same flavour and opposite charge passing the impact parameter selection. 
The first control sample is obtained by requiring that the two loose leptons, which do not form the $\text{\PZ}_{1}$ candidate, do not pass the final identification and isolation criteria (2P2F region). 
The second control sample is obtained by requiring that one of the $4\ell$ does not pass the final identification and isolation criteria, and the other three leptons do (3P1F region). 

The expected number of events from reducible backgrounds in the SR can be written as described in Ref.~\cite{Sirunyan:2017exp}:

\begin{equation}
  \label{eq:fakerate2}
  N^{\text{Z+X}}_{\mathrm{SR}} = \Sigma \frac{f_{i}}{(1 - f_{i})} (N_{3\text{P}1\text{F}} - N^{\text{bkg}}_{3\text{P}1\text{F}} - N^{\PZ\PZ}_{3\text{P}1\text{F}} ) + \Sigma \frac{f_{i}}{(1 - f_{i})} \frac{f_{j}}{(1 - f_{j})} N_{2\text{P}2\text{F}},
\end{equation}

where $N_{2\text{P}2\text{F}}$ is the observed number of events in the first control sample, $N_{3\text{P}1\text{F}}$ is the observed number of events in the second control sample, $N^{\text{bkg}}_{3\text{P}1\text{F}}$ is the contamination of 2P2F in the 3P1F CR, $N^{\PZ\PZ}_{3\text{P}1\text{F}}$ is the number of background events which have 4 real leptons and 2 jets estimated from MC simulation, and $f_i$ and $f_j$ are the estimated misidentification probabilities of the two loose leptons.

The dominant systematic uncertainty in the reducible background estimation arises mainly from the limitated number of events both in the CR and in the region where the misidentification rates are computed. Uncertainties that arise from the difference in the composition of the sample from which the misidentification rate is computed are also taken into account.

\section{Multivariate analysis}
\label{sec:bdt}

A boosted decision tree (BDT) is used to further discriminate signal from backgrounds. The BDT is implemented in the TMVA tool of the ROOT analysis package \cite{Hocker:2007ht}, and trained using simulated events in the 4$\ell$ SR; training is done using the gluon-gluon \HH process as signal and all the other processes as background.

An optimal set of input variables was chosen maximizing the area under the receiver operating characteristic curve for each resulting discriminator. The optimized set contains ten variables, namely: the \pt of each of the $4\ell$, the $\Delta R$ between the reconstructed $\PH\to \PZ\PZ^{*} \to 4\ell$ and $\PH\to \PQb\PAQb$ systems, the two $\PQb$-tagging scores, the \pt of each of the two jets with the highest value of $\PQb$-tagging score and their invariant mass. The most important variables for the BDT training are the $\PQb$-tagging scores of the two selected jets, the invariant mass of the dijet system, and the $\Delta R$ between the reconstructed $\PH\to \PZ\PZ^{*} \to 4\ell$ and $\PH\to \PQb\PAQb$ systems.

\begin{figure}[htb!]
  \centering
  \includegraphics[width=0.48\columnwidth]{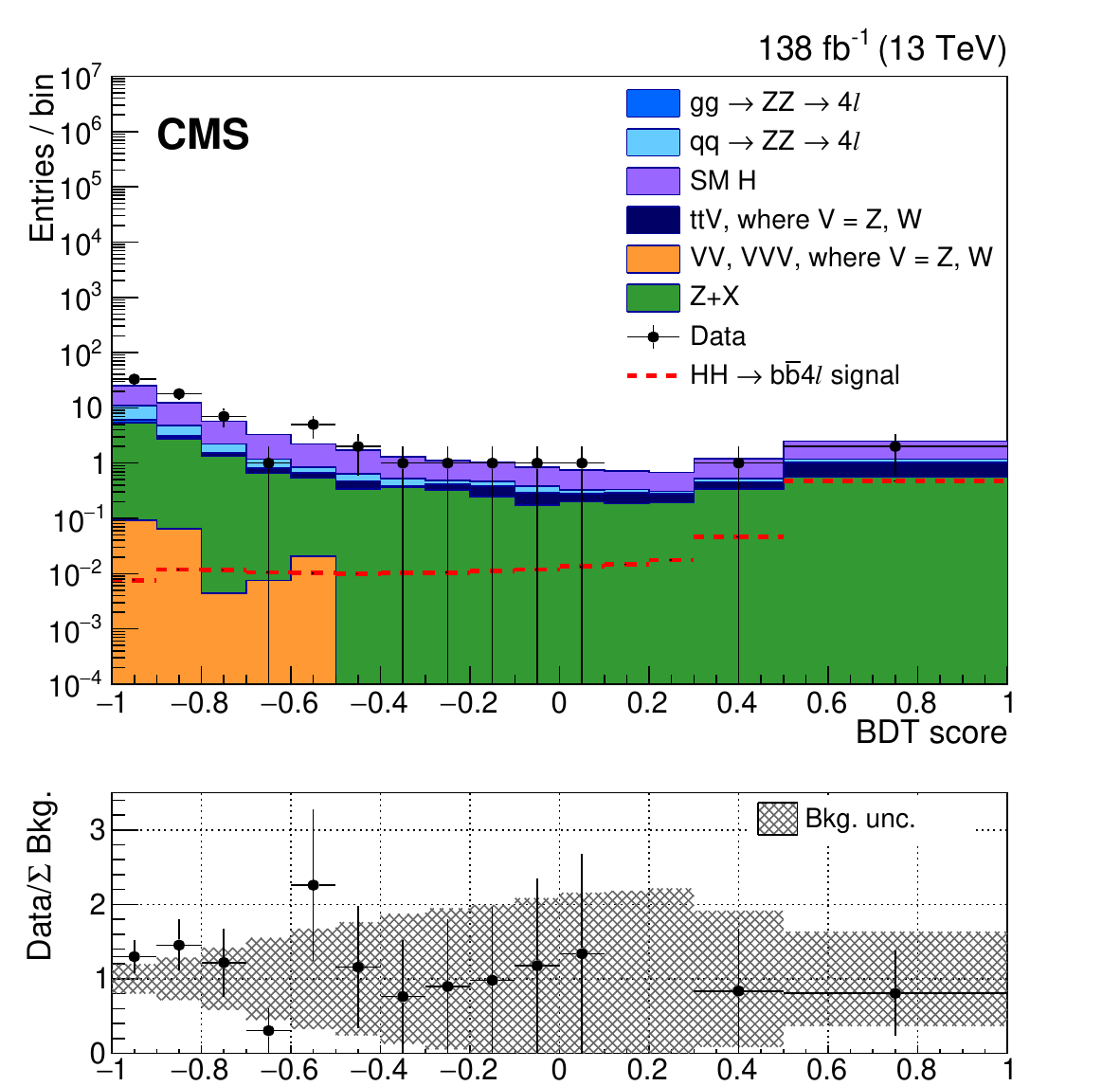}
   \caption{Inclusive BDT distributions for simulated signal, estimated background components, and data for the combination of the three different final states (4$\PGm$, 4$\Pe$, and $2\Pe2\PGm$) and data-taking years. In the ratio plot, the grey dashed shades represent the total uncertainties on the estimated background.}
  \label{fig:bdt}
\end{figure}

To take into account the possible differences due to data-taking conditions and final states, the BDT is trained separately for each data-taking year and for each of the leptonic final states (4$\mu$, 4$\Pe$, and 2e$2\mu$), for a total of nine independent discriminators. Detailed studies were performed on the discrimination ranking, variable correlations, and possible discriminator overtraining. Figure \ref{fig:bdt} shows the inclusive BDT discriminator distribution for simulated signal, estimated background components, and data: different data-taking years and leptonic final states are combined together (4$\PGm$, 4$\Pe$, and 2$\Pe$2$\PGm$).
Signal and backgrounds normalisation results from the fit procedure described in Section \ref{sec:results}.
The binning of the BDT output distribution is chosen in order to have (almost) the same statistical uncertainties in the last 4--5 bins, which is around 10\% and which ensures stable results under statistical fluctuations of the most sensitive bins.

The output of the BDT is used in a statistical analysis as the input variable of a maximum likelihood fit to set constraints on both the signal strength and the $\kappa_\lambda$ coupling modifier.

\section{Systematic uncertainties}
\label{sec:systematics}

Systematic uncertainties are modelled in a maximum likelihood fit as nuisance parameters.
Two types of nuisance parameters are considered: \emph{normalization} uncertainties, describing possible modification to the event yields, and \emph{shape} uncertainties, accounting also for variations of the shape of distributions.
The former are assigned a log-normal probability density function, while the latter are included as template shape variations and taken into account via continuous template morphing~\cite{CombineManual}.

The experimental uncertainties considered for the measurement arise from different sources. The uncertainty in the integrated luminosity is 1.2\% for 2016 \cite{CMS-LUM-17-003}, 2.3\% for 2017 \cite{CMS-PAS-LUM-17-004}, and 2.5\% for 2018 \cite{CMS-PAS-LUM-18-002}. The uncertainty in the lepton identification and reconstruction efficiency ranges $1.0--15.5 \%$ on the overall event yield for the different final states. Muon uncertainties are always below 3\%, while electron uncertainties surpass 10\% for electrons with $\pt < 20 \GeV$ and $|\eta| > 1.5$.
The uncertainties in the jet energy scale (JES) and jet energy resolution (JER) are accounted for by changing the jet \pt by one standard deviation for each of the several uncorrelated sources considered \cite{JINST6}. The effects on the signal acceptance are also accounted for.
Uncertainties in the scale factors are computed and applied as a function of the jet \pt, $\eta$, and hadron flavour.
The above uncertainties apply equally to all simulated signal and background samples.
The JES, JER, and $\PQb$-tagging scale factor systematics are considered shape uncertainties.
The JES, together with the statistical uncertainties in the yield of the last bin of the BDT, are the systematic effects with the largest impact on the final result. 
All the other sources are considered normalization uncertainties.
The experimental uncertainties originating from the reducible background estimation, described in Section~\ref{sec:redBkg}, are also considered. The main contribution comes from the mismatch in the composition of backgrounds between the samples where the misidentification probability is derived and those where it is applied. The summary of the experimental systematic uncertainties is reported in Table~\ref{tab:systexpunc}. 

Theoretical uncertainties arise from the choice of the PDF set, the uncertainty in $\alpha_S$, and the renormalization and factorization scales \cite{Cepeda:2019klc}: they affect both signal and background processes. For the \HH signal, in addition to the uncertainty sources just described, an uncertainty caused by missing finite $m_{top}$ effects \cite{Baglio:2020wgt} contributes, too. 
No additional uncertainties on \PH production without real \PQb jets are applied since they are negligible compared to all the other uncertainties involved.
An additional uncertainty of 10\% for the $\Pg\Pg\to\PZ\PZ^{*}$ prediction and of 0.1\% for the $\PQq\PQq\to\PZ\PZ^{*}$ prediction is considered to account for the uncertainty in the applied \textit{K}-factors, described in Section~\ref{sec:datasets}.

The uncertainty in the luminosity is treated as partially correlated across data-taking periods. All the other experimental uncertainties are considered uncorrelated: they are estimated for each year independently and therefore their dominating statistical uncertainty is uncorrelated. All theoretical uncertainties are considered correlated among the three data-taking years.
A summary of the theory systematic uncertainties is reported in Table~\ref{tab:systtheorunc}.

\begin{table}[ht]
  \centering
  \topcaption{Summary of experimental systematic uncertainties. The effect of $\PQb$-tagging SF, JER, and JES is propagated to the shape of the BDT distribution.}
      \begin{tabular}{lccc}
      \hline
      \multicolumn{4}{c}{Experimental uncertainties} \\
      \hline
       Source                             & 2016              & 2017              &  2018             \\
      \hline
       Integrated luminosity                       &  1.2\%          &  2.3\%          &  2.5\%          \\
       Lepton reco./ident. eff.         &  1.6--15.5\%      &  1.1--12.1\%   &  1.0--11.0\%   \\
       $\PQb$-tagging SF                     &  shape &  shape &  shape \\
       JES                              &  shape &  shape &  shape \\
       JER                              &  shape &  shape &  shape \\
       \PZ{}+X bkg. uncertainties              &  30--41\% &  30--38\% &  30--37\% \\
      \hline 
      \end{tabular}
\label{tab:systexpunc}
\end{table}

\begin{table}[ht]
  \centering
  \topcaption{Summary of the theory systematic uncertainties considered in this analysis.}
      \begin{tabular}{lc}
           \hline
     \multicolumn{2}{c}{Theory uncertainties} \\
     \hline
     \hline
     \multicolumn{2}{c}{Branching fractions} \\
     \hline
     $\mathcal{B}(\PH \to \PQb\PAQb)$       &  1.2\% \\
     $\mathcal{B}(\PH \to \PZ\PZ)$          &  1.5\% \\     
     \hline  
     \multicolumn{2}{c}{PDF set and $\alpS$} \\
     \hline
       \HH                                  &  3.0\% \\
       $\HH$ ($m_{\text{top}}$ effects)            &  $^{+4.0}_{-18.0}$\% \\ 
       $\Pg\Pg\PH$ (PDF set)                &  1.9\% \\
       $\Pg\Pg\PH$ ($\alpS$)                &  2.59--2.62\% \\
       VBF$\PH$                             &  2.1\% \\
       $\PZ\PH$                             &  1.6\% \\
       $\PW\PH$                             &  1.3\% \\
       $\bbH$                               &  3.2\% \\
       $\ttbar\PH$                          &  3.0\% \\
       $\PQq\PQq\PZ\PZ$                     &  3.1--3.4\% \\
       $\ttbar\PZ$                          &  7--14\% \\
       VVV                                  &  2--17\% \\
       $\Pg\Pg\PZ\PZ$                       &  3.2\%  \\
     \hline
     \multicolumn{2}{c}{Scale uncertainties} \\  
     \hline
       \HH                                  &  2.2--5\% \\
       $\Pg\Pg\PH$                          &  4.27--6.49\% \\
       VBF$\PH$                             &  0.3--0.4\% \\
       $\PZ\PH$                             &  2.7--3.5\% \\
       $\ttbar\PH$                          &  6.0--9.2\% \\
       $\PQq\PQq\PZ\PZ$                     &  3.2--4.2\% \\
       $\ttbar\PZ$                          &  2--3\% \\
       VVV                                  &  3\% \\
       $\Pg\Pg\PZ\PZ$                       &  4.6--6.7\% \\
     \hline
     \multicolumn{2}{c}{\textit{K}-factors} \\
     \hline
       $\PQq\PQq\PZ\PZ$                                                  &  0.1\% \\
       $\Pg\Pg\PZ\PZ$                                                    &  10.0\% \\
     \hline 
      \end{tabular}
\label{tab:systtheorunc}
\end{table}

This analysis is still dominated by statistical uncertainties, thus the impact of the systematic uncertainties is overall negligible.

\clearpage

\section{Results}
\label{sec:results}

No significant excess of events is observed over the expected backgrounds. Upper limits are set on the production cross section of \HH times the branching fraction ${\mathcal{B}(\PH\PH\to4\ell \PQb\PAQb)}$, using the modified frequentist approach (\CLs), taking the profile likelihood
as a test statistic~\cite{Read:2002hq,CLS2,CLSA,CMS-NOTE-2011-005} in the asymptotic approximation. The limits are compared to the theoretical predictions assuming SM branching fractions for the \PH decays.

The results are extracted with a binned maximum likelihood fit to the nine BDT distributions in data. Shape templates for the BDT score for signal are obtained from simulated samples and for backgrounds from both simulated samples and data, as described in Section~\ref{sec:bdt}.
Systematic uncertainties are treated as described in Section~\ref{sec:systematics}.

\begin{figure}[ht!]
  \centering
  \includegraphics[width=0.48\textwidth]{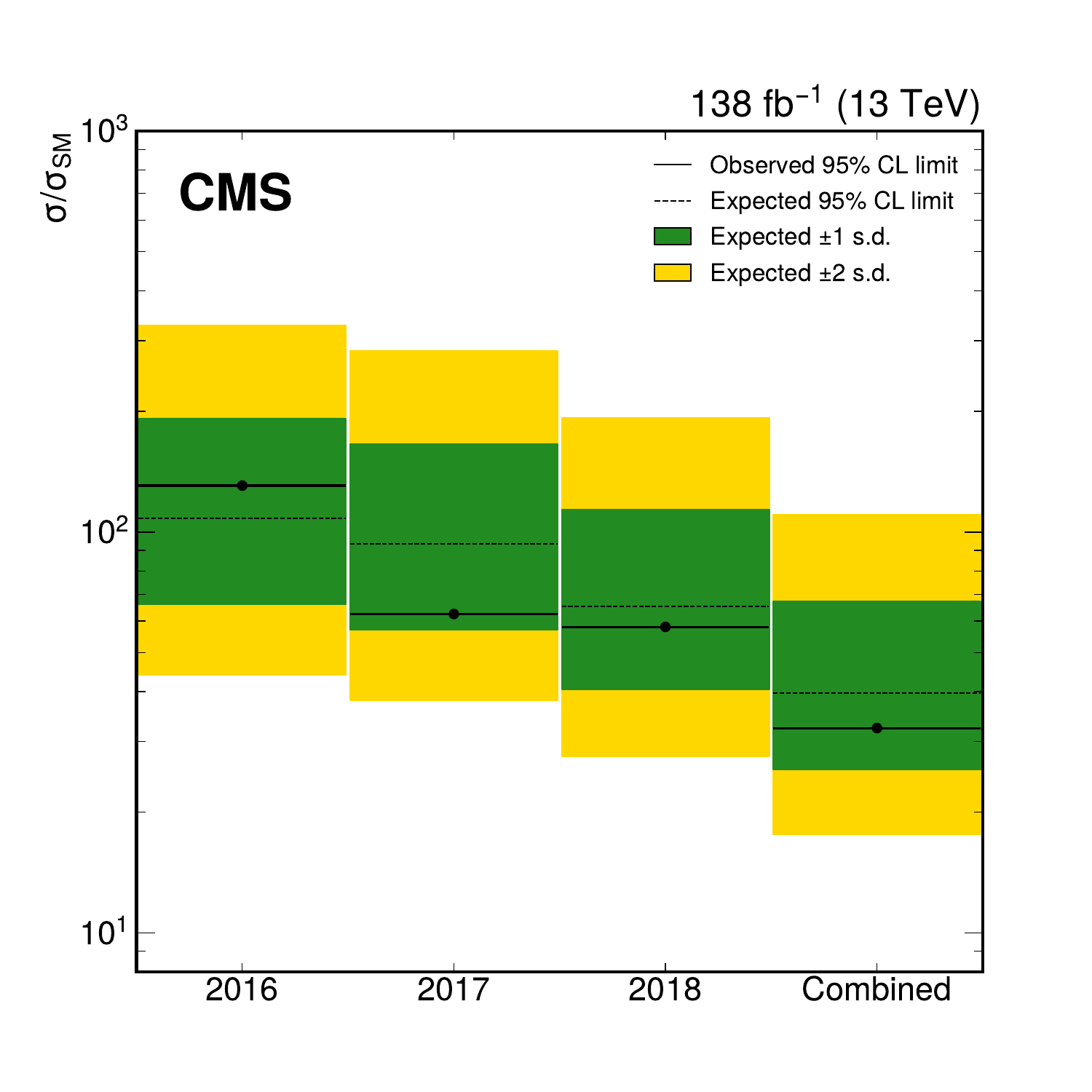}
  \includegraphics[width=0.48\textwidth]{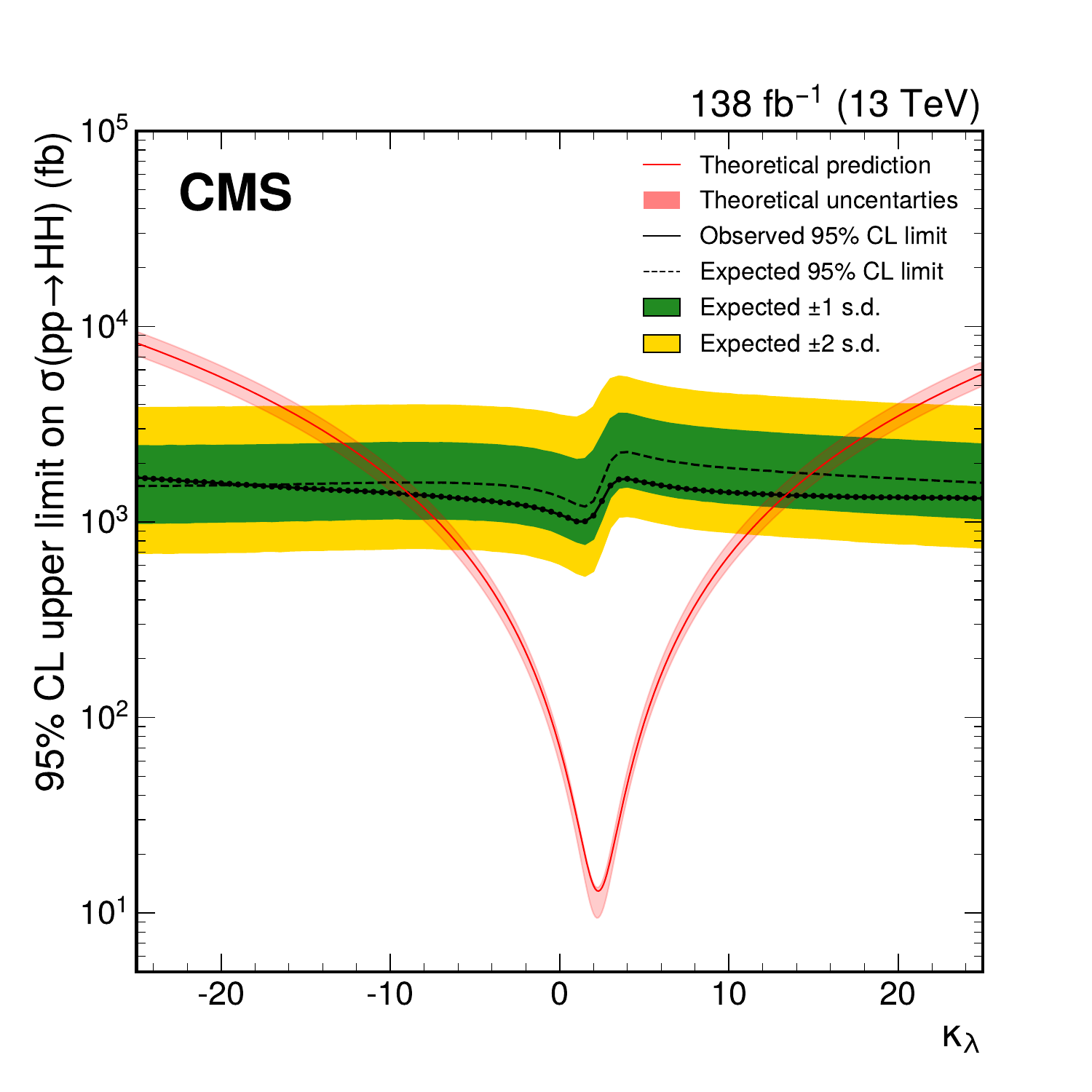}
  \caption{Left: upper limit on the signal strength at 95\% \CL for each year and for their combination. Right: expected and observed 95\% \CL upper limits on the SM-like \HH production cross section obtained for different values of  $\kappa_\lambda$. The green and yellow bands represent, respectively, the one and two standard deviation (s.d.) quantiles around the expected limit. The red curve is the \HH theoretical prediction cross section as a function of $\kappa_\lambda$.}
  \label{fig:klambdascan}
\end{figure}

The observed (expected) upper limit on the signal strength modifier $\mu$, defined as the ratio of the \HH fraction in the $4\ell\PQb\PAQb$ final state to the SM expectation, is 32.4 (39.6) at 95\% \CL, as illustrated in Fig.~\ref{fig:klambdascan}~(left).

Upper limits are also set for different hypotheses of anomalous values of the \PH self-coupling $\kappa_\lambda$, assuming all the other couplings~\cite{Dall'Osso:2015aia} are equal to their SM values. The result is shown in Fig.~\ref{fig:klambdascan}~(right), where the exclusion is compared to the theoretical prediction for the \HH cross section (red line). 
The analysis constrains $\kappa_{\lambda}$ to be within the observed (expected) range $-8.8\,(-9.8) < \kappa_{\lambda} < 13.4\,(15.0)$ at 95\% \CL.

The limits set by this analysis are obtained with a data set which is independent from those used for the other channels~\cite{Khachatryan:2016sey, Khachatryan:2015year,Khachatryan:2016sey, Khachatryan:2015year}.

The results presented in this section are provided in a tabulated form in the HEPDATA record~\cite{hepData} for this analysis.

\clearpage

\section{Summary}

The first search for nonresonant Higgs boson (\PH) pair production in the four lepton $\PQb\PAQb$ final state is presented. 
This search uses a data sample collected in proton-proton collisions at $\sqrt{s} = 13~\text{TeV}$ by the CMS detector during the years 2016--2018 corresponding to an integrated luminosity of 138~\fbinv. 
The results are found to be compatible with the standard model (SM) expectations. An observed (expected) upper limit on the HH production cross section is set at 32.4 (39.6) times the SM expected rate at the 95\% confidence level.
Possible modifications of the \PH trilinear coupling $\lambda_{\PH\PH\PH}$ with respect to the standard model (SM) value are investigated.
The coupling modifier $\kappa_{\lambda}$, defined as $\lambda_{\PH\PH\PH}$ divided by its SM prediction, is constrained to be within the observed (expected) range $-8.8\,(-9.8) < \kappa_{\lambda} < 13.4\,(15.0)$ at 95\% confidence level.

\begin{acknowledgments}
We congratulate our colleagues in the CERN accelerator departments for the excellent performance of the LHC and thank the technical and administrative staffs at CERN and at other CMS institutes for their contributions to the success of the CMS effort. In addition, we gratefully acknowledge the computing centres and personnel of the Worldwide LHC Computing Grid and other centres for delivering so effectively the computing infrastructure essential to our analyses. Finally, we acknowledge the enduring support for the construction and operation of the LHC, the CMS detector, and the supporting computing infrastructure provided by the following funding agencies: BMBWF and FWF (Austria); FNRS and FWO (Belgium); CNPq, CAPES, FAPERJ, FAPERGS, and FAPESP (Brazil); MES and BNSF (Bulgaria); CERN; CAS, MoST, and NSFC (China); MINCIENCIAS (Colombia); MSES and CSF (Croatia); RIF (Cyprus); SENESCYT (Ecuador); MoER, ERC PUT and ERDF (Estonia); Academy of Finland, MEC, and HIP (Finland); CEA and CNRS/IN2P3 (France); BMBF, DFG, and HGF (Germany); GSRI (Greece); NKFIH (Hungary); DAE and DST (India); IPM (Iran); SFI (Ireland); INFN (Italy); MSIP and NRF (Republic of Korea); MES (Latvia); LAS (Lithuania); MOE and UM (Malaysia); BUAP, CINVESTAV, CONACYT, LNS, SEP, and UASLP-FAI (Mexico); MOS (Montenegro); MBIE (New Zealand); PAEC (Pakistan); MES and NSC (Poland); FCT (Portugal); MESTD (Serbia); MCIN/AEI and PCTI (Spain); MOSTR (Sri Lanka); Swiss Funding Agencies (Switzerland); MST (Taipei); MHESI and NSTDA (Thailand); TUBITAK and TENMAK (Turkey); NASU (Ukraine); STFC (United Kingdom); DOE and NSF (USA).
 
 \hyphenation{Rachada-pisek} Individuals have received support from the Marie-Curie programme and the European Research Council and Horizon 2020 Grant, contract Nos.\ 675440, 724704, 752730, 758316, 765710, 824093, 884104, and COST Action CA16108 (European Union); the Leventis Foundation; the Alfred P.\ Sloan Foundation; the Alexander von Humboldt Foundation; the Belgian Federal Science Policy Office; the Fonds pour la Formation \`a la Recherche dans l'Industrie et dans l'Agriculture (FRIA-Belgium); the Agentschap voor Innovatie door Wetenschap en Technologie (IWT-Belgium); the F.R.S.-FNRS and FWO (Belgium) under the ``Excellence of Science -- EOS" -- be.h project n.\ 30820817; the Beijing Municipal Science \& Technology Commission, No. Z191100007219010; the Ministry of Education, Youth and Sports (MEYS) of the Czech Republic; the Hellenic Foundation for Research and Innovation (HFRI), Project Number 2288 (Greece); the Deutsche Forschungsgemeinschaft (DFG), under Germany's Excellence Strategy -- EXC 2121 ``Quantum Universe" -- 390833306, and under project number 400140256 - GRK2497; the Hungarian Academy of Sciences, the New National Excellence Program - \'UNKP, the NKFIH research grants K 124845, K 124850, K 128713, K 128786, K 129058, K 131991, K 133046, K 138136, K 143460, K 143477, 2020-2.2.1-ED-2021-00181, and TKP2021-NKTA-64 (Hungary); the Council of Science and Industrial Research, India; the Latvian Council of Science; the Ministry of Education and Science, project no. 2022/WK/14, and the National Science Center, contracts Opus 2021/41/B/ST2/01369 and 2021/43/B/ST2/01552 (Poland); the Funda\c{c}\~ao para a Ci\^encia e a Tecnologia, grant CEECIND/01334/2018 (Portugal); the National Priorities Research Program by Qatar National Research Fund; MCIN/AEI/10.13039/501100011033, ERDF ``a way of making Europe", and the Programa Estatal de Fomento de la Investigaci{\'o}n Cient{\'i}fica y T{\'e}cnica de Excelencia Mar\'{\i}a de Maeztu, grant MDM-2017-0765 and Programa Severo Ochoa del Principado de Asturias (Spain); the Chulalongkorn Academic into Its 2nd Century Project Advancement Project, and the National Science, Research and Innovation Fund via the Program Management Unit for Human Resources \& Institutional Development, Research and Innovation, grant B05F650021 (Thailand); the Kavli Foundation; the Nvidia Corporation; the SuperMicro Corporation; the Welch Foundation, contract C-1845; and the Weston Havens Foundation (USA).  
\end{acknowledgments}

\bibliography{auto_generated}

\providecommand{\href}[2]{#2}\begingroup\raggedright\begin{thebibliography}{10}%
\makeatletter
\providecommand{\hrefCMSnoop }[0]{\@secondoftwo}%
\makeatother
\providecommand{\doi}{\texttt{doi:}\begingroup \urlstyle{tt}\Url}

\bibitem{Aad:2012tfa}
\hrefCMSnoop {}{{ATLAS Collaboration}, ``{Observation of a new particle in the
  search for the standard model Higgs boson with the ATLAS detector at the
  LHC}'',} \textit{ Phys. Lett. B} \textbf{ 716} (2012) 1,
  \href{http://dx.doi.org/10.1016/j.physletb.2012.08.020}{\doi{10.1016/j.physletb.2012.08.020}},
\href{http://www.arXiv.org/abs/1207.7214}{\texttt{arXiv:1207.7214}}.

\bibitem{Chatrchyan:2012ufa}
\hrefCMSnoop {}{{CMS Collaboration}, ``{Observation of a new boson with mass
  near 125 GeV in pp collisions at $\sqrt{s}$ = 7 and 8 TeV}'',} \textit{ JHEP}
  \textbf{ 06} (2013) 081. 117,
  \href{http://dx.doi.org/10.1007/JHEP06(2013)081}{\doi{10.1007/JHEP06(2013)081}},
\href{http://www.arXiv.org/abs/1303.4571}{\texttt{arXiv:1303.4571}}.

\bibitem{Aad:2015zhl}
\hrefCMSnoop {}{{ATLAS and CMS Collaborations}, ``{Combined measurement of the
  Higgs boson mass in pp collisions at $\sqrt{s}=7$ and 8 TeV with the ATLAS
  and CMS experiments}'',} \textit{ Phys. Rev. Lett.} \textbf{ 114} (2015)
  191803,
  \href{http://dx.doi.org/10.1103/PhysRevLett.114.191803}{\doi{10.1103/PhysRevLett.114.191803}},
\href{http://www.arXiv.org/abs/1503.07589}{\texttt{arXiv:1503.07589}}.

\bibitem{Sirunyan:2017exp}
\hrefCMSnoop {}{{CMS Collaboration}, ``{Measurements of properties of the Higgs
  boson decaying into the four-lepton final state in pp collisions at $
  \sqrt{s}=13 $ TeV}'',} \textit{ JHEP} \textbf{ 11} (2017) 047,
  \href{http://dx.doi.org/10.1007/JHEP11(2017)047}{\doi{10.1007/JHEP11(2017)047}},
\href{http://www.arXiv.org/abs/1706.09936}{\texttt{arXiv:1706.09936}}.

\bibitem{CMS:2018lkl}
\hrefCMSnoop {}{{CMS Collaboration}, ``{Combined measurements of Higgs boson
  couplings in proton--proton collisions at $\sqrt{s}=13\,\text {Te}\text {V}
  $}'',} \textit{ Eur. Phys. J. C} \textbf{ 79} (2019) 421,
  \href{http://dx.doi.org/10.1140/epjc/s10052-019-6909-y}{\doi{10.1140/epjc/s10052-019-6909-y}},
  \href{http://www.arXiv.org/abs/1809.10733}{\texttt{arXiv:1809.10733}}.

\bibitem{Baglio:2012np}
J.~Baglio\hrefCMSnoop {}{ {et~al.}, ``{The measurement of the Higgs
  self-coupling at the LHC: theoretical status}'',} \textit{ JHEP} \textbf{ 04}
  (2013) 151,
  \href{http://dx.doi.org/10.1007/JHEP04(2013)151}{\doi{10.1007/JHEP04(2013)151}},
\href{http://www.arXiv.org/abs/1212.5581}{\texttt{arXiv:1212.5581}}.

\bibitem{Aad:2014yja}
\hrefCMSnoop {}{{ATLAS Collaboration}, ``{Search for Higgs boson pair
  production in the $\gamma\gamma \PQb\PAQb$ final state using pp collision
  data at $\sqrt{s}=8$ TeV from the ATLAS detector}'',} \textit{ Phys. Rev.
  Lett.} \textbf{ 114} (2015) 081802,
  \href{http://dx.doi.org/10.1103/PhysRevLett.114.081802}{\doi{10.1103/PhysRevLett.114.081802}},
\href{http://www.arXiv.org/abs/1406.5053}{\texttt{arXiv:1406.5053}}.

\bibitem{Aad:2015uka}
\hrefCMSnoop {}{{ATLAS Collaboration}, ``{Search for Higgs boson pair
  production in the $\PQb\PAQb\PQb\PAQb$ final state from pp collisions at
  $\sqrt{s} = 8$ TeV with the ATLAS detector}'',} \textit{ Eur. Phys. J. C}
  \textbf{ 75} (2015) 412,
  \href{http://dx.doi.org/10.1140/epjc/s10052-015-3628-x}{\doi{10.1140/epjc/s10052-015-3628-x}},
\href{http://www.arXiv.org/abs/1506.00285}{\texttt{arXiv:1506.00285}}.

\bibitem{Aad:2015xja}
\hrefCMSnoop {}{{ATLAS Collaboration}, ``{Searches for Higgs boson pair
  production in the $\PH\PH\to\PQb\PQb\tau\tau, \gamma\gamma \PW\PW^*,
  \gamma\gamma \PQb\PQb, \PQb\PQb\PQb\PQb$ channels with the ATLAS
  detector}'',} \textit{ Phys. Rev. D} \textbf{ 92} (2015) 092004,
  \href{http://dx.doi.org/10.1103/PhysRevD.92.092004}{\doi{10.1103/PhysRevD.92.092004}},
\href{http://www.arXiv.org/abs/1509.04670}{\texttt{arXiv:1509.04670}}.

\bibitem{Aaboud:2016xco}
\hrefCMSnoop {}{{ATLAS Collaboration}, ``Search for pair production of {H}iggs
  bosons in the $\bbbar\bbbar$ final state using proton--proton collisions at
  $\sqrt{s} = 13$\,{TeV} with the {ATLAS} detector'',} \textit{ Phys. Rev. D}
  \textbf{ 94} (2016) 052002,
  \href{http://dx.doi.org/10.1103/PhysRevD.94.052002}{\doi{10.1103/PhysRevD.94.052002}},
\href{http://www.arXiv.org/abs/1606.04782}{\texttt{arXiv:1606.04782}}.

\bibitem{Khachatryan:2016sey}
\hrefCMSnoop {}{{CMS Collaboration}, ``{Search for two Higgs bosons in final
  states containing two photons and two bottom quarks in proton-proton
  collisions at 8 TeV}'',} \textit{ Phys. Rev. D} \textbf{ 94} (2016) 052012,
  \href{http://dx.doi.org/10.1103/PhysRevD.94.052012}{\doi{10.1103/PhysRevD.94.052012}},
\href{http://www.arXiv.org/abs/1603.06896}{\texttt{arXiv:1603.06896}}.

\bibitem{Khachatryan:2015year}
\hrefCMSnoop {}{{CMS Collaboration}, ``{Search for resonant pair production of
  Higgs bosons decaying to two bottom quark-antiquark pairs in proton-proton
  collisions at 8 TeV}'',} \textit{ Phys. Lett. B} \textbf{ 749} (2015) 560,
  \href{http://dx.doi.org/10.1016/j.physletb.2015.08.047}{\doi{10.1016/j.physletb.2015.08.047}},
\href{http://www.arXiv.org/abs/1503.04114}{\texttt{arXiv:1503.04114}}.

\bibitem{Sirunyan:2017isc}
\hrefCMSnoop {}{{CMS Collaboration}, ``{Search for a massive resonance decaying
  to a pair of Higgs bosons in the four b quark final state in proton-proton
  collisions at $\sqrt{s}=13$\,{TeV}}'',} \textit{ Phys. Lett. B} \textbf{ 781}
  (2018) 244,
  \href{http://dx.doi.org/10.1016/j.physletb.2018.03.084}{\doi{10.1016/j.physletb.2018.03.084}},
\href{http://www.arXiv.org/abs/1710.04960}{\texttt{arXiv:1710.04960}}.

\bibitem{Sirunyan:2017tqo}
\hrefCMSnoop {}{{CMS Collaboration}, ``Search for higgs boson pair production
  in the $\bbbar\tau\tau$ final state in proton-proton collisions at
  $\sqrt{s}=8$\,{TeV}'',} \textit{ Phys. Rev. D} \textbf{ 96} (2017) 072004,
  \href{http://dx.doi.org/10.1103/PhysRevD.96.072004}{\doi{10.1103/PhysRevD.96.072004}},
\href{http://www.arXiv.org/abs/1707.00350}{\texttt{arXiv:1707.00350}}.

\bibitem{Sirunyan:2017djm}
\hrefCMSnoop {}{{CMS Collaboration}, ``{Search for Higgs boson pair production
  in events with two bottom quarks and two tau leptons in proton--proton
  collisions at $\sqrt s$ = 13~TeV}'',} \textit{ Phys. Lett. B} \textbf{ 778}
  (2018) 101,
  \href{http://dx.doi.org/10.1016/j.physletb.2018.01.001}{\doi{10.1016/j.physletb.2018.01.001}},
  \href{http://www.arXiv.org/abs/1707.02909}{\texttt{arXiv:1707.02909}}.

\bibitem{Sirunyan:2017guj}
\hrefCMSnoop {}{{CMS Collaboration}, ``Search for resonant and nonresonant
  {H}iggs boson pair production in the $\bbbar\ell\nu\ell\nu$ final state in
  proton-proton collisions at $ \sqrt{s}=13$\,{TeV}'',} \textit{ JHEP} \textbf{
  01} (2018) 054,
  \href{http://dx.doi.org/10.1007/JHEP01(2018)054}{\doi{10.1007/JHEP01(2018)054}},
\href{http://www.arXiv.org/abs/1708.04188}{\texttt{arXiv:1708.04188}}.

\bibitem{CMS:2020gsy}
\href {http://cds.cern.ch/record/2706103}{{CMS Collaboration}, ``{Combined
  Higgs boson production and decay measurements with up to 137 $\fbinv$ of
  proton-proton collision data at $\sqrt{s}$ = 13 TeV}'',} CMS Physics Analysis
  Summary CMS-PAS-HIG-19-005, 2020.

\bibitem{ATLAScomb}
\hrefCMSnoop {}{{ATLAS Collaboration}, ``{Combination of searches for Higgs
  boson pairs in $\Pp\Pp$ collisions at $\sqrt{s} = $~13~TeV with the ATLAS
  detector}'',} \textit{ Phys. Lett. B} \textbf{ 800} (2020) 135103,
  \href{http://dx.doi.org/10.1016/j.physletb.2019.135103}{\doi{10.1016/j.physletb.2019.135103}},
  \href{http://www.arXiv.org/abs/1906.02025}{\texttt{arXiv:1906.02025}}.

\bibitem{CMScomb}
\hrefCMSnoop {}{{CMS Collaboration}, ``{Combination of searches for Higgs boson
  pair production in proton-proton collisions at $\sqrt{s} = $ 13 TeV}'',}
  \textit{ Phys. Rev. Lett.} \textbf{ 122} (2019) 121803,
  \href{http://dx.doi.org/10.1103/PhysRevLett.122.121803}{\doi{10.1103/PhysRevLett.122.121803}},
  \href{http://www.arXiv.org/abs/1811.09689}{\texttt{arXiv:1811.09689}}.

\bibitem{Baglio:2020wgt}
J.~Baglio\hrefCMSnoop {}{ {et~al.}, ``{$\Pg\Pg \to \PH\PH$ : Combined
  uncertainties}'',} \textit{ Phys. Rev. D} \textbf{ 103} (2021) 056002,
  \href{http://dx.doi.org/10.1103/PhysRevD.103.056002}{\doi{10.1103/PhysRevD.103.056002}},
  \href{http://www.arXiv.org/abs/2008.11626}{\texttt{arXiv:2008.11626}}.

\bibitem{Heinrich:2017kxx}
G.~Heinrich\hrefCMSnoop {}{ {et~al.}, ``{NLO predictions for Higgs boson pair
  production with full top quark mass dependence matched to parton showers}'',}
  \textit{ JHEP} \textbf{ 08} (2017) 088,
  \href{http://dx.doi.org/10.1007/JHEP08(2017)088}{\doi{10.1007/JHEP08(2017)088}},
\href{http://www.arXiv.org/abs/1703.09252}{\texttt{arXiv:1703.09252}}.

\bibitem{Buchalla:2018yce}
G.~Buchalla\hrefCMSnoop {}{ {et~al.}, ``{Higgs boson pair production in
  non-linear Effective Field Theory with full $m_\text{t}$-dependence at NLO
  QCD}'',} \textit{ JHEP} \textbf{ 09} (2018) 057,
  \href{http://dx.doi.org/10.1007/JHEP09(2018)057}{\doi{10.1007/JHEP09(2018)057}},
\href{http://www.arXiv.org/abs/1806.05162}{\texttt{arXiv:1806.05162}}.

\bibitem{Goertz:2014qta}
\hrefCMSnoop {}{F.~Goertz, A.~Papaefstathiou, L.~L. Yang, and J.~Zurita,
  ``{Higgs boson pair production in the D=6 extension of the SM}'',} \textit{
  JHEP} \textbf{ 04} (2015) 167,
  \href{http://dx.doi.org/10.1007/JHEP04(2015)167}{\doi{10.1007/JHEP04(2015)167}},
\href{http://www.arXiv.org/abs/1410.3471}{\texttt{arXiv:1410.3471}}.

\bibitem{Chatrchyan:2008zzk}
\hrefCMSnoop {}{{CMS Collaboration}, ``The {CMS} experiment at the {CERN}
  {LHC}'',} \textit{ JINST} \textbf{ 3} (2008) S08004,
  \href{http://dx.doi.org/10.1088/1748-0221/3/08/S08004}{\doi{10.1088/1748-0221/3/08/S08004}}.

\bibitem{h4lFullRun2}
\hrefCMSnoop {}{{CMS Collaboration}, ``Measurements of production cross
  sections of the {H}iggs boson in the four-lepton final state in proton-proton
  collisions at {$\sqrt{s} = 13 \TeV$}'',} \textit{ Eur. Phys. J. C} \textbf{
  81} (2021) 488,
  \href{http://dx.doi.org/10.1140/epjc/s10052-021-09200-x}{\doi{10.1140/epjc/s10052-021-09200-x}},
  \href{http://www.arXiv.org/abs/2103.04956}{\texttt{arXiv:2103.04956}}.

\bibitem{Alioli:2008gx}
\hrefCMSnoop {}{S.~Alioli, P.~Nason, C.~Oleari, and E.~Re, ``{NLO vector-boson
  production matched with shower in POWHEG}'',} \textit{ JHEP} \textbf{ 07}
  (2008) 060,
  \href{http://dx.doi.org/10.1088/1126-6708/2008/07/060}{\doi{10.1088/1126-6708/2008/07/060}},
\href{http://www.arXiv.org/abs/0805.4802}{\texttt{arXiv:0805.4802}}.

\bibitem{Nason:2004rx}
\hrefCMSnoop {}{P.~Nason, ``{A new method for combining NLO QCD with shower
  Monte Carlo algorithms}'',} \textit{ JHEP} \textbf{ 11} (2004) 040,
  \href{http://dx.doi.org/10.1088/1126-6708/2004/11/040}{\doi{10.1088/1126-6708/2004/11/040}},
\href{http://www.arXiv.org/abs/hep-ph/0409146}{\texttt{arXiv:hep-ph/0409146}}.

\bibitem{Frixione:2007vw}
\hrefCMSnoop {}{S.~Frixione, P.~Nason, and C.~Oleari, ``{Matching NLO QCD
  computations with parton shower simulations: the POWHEG method}'',} \textit{
  JHEP} \textbf{ 11} (2007) 070,
  \href{http://dx.doi.org/10.1088/1126-6708/2007/11/070}{\doi{10.1088/1126-6708/2007/11/070}},
\href{http://www.arXiv.org/abs/0709.2092}{\texttt{arXiv:0709.2092}}.

\bibitem{Nason:2009ai}
\hrefCMSnoop {}{P.~Nason and C.~Oleari, ``{NLO Higgs boson production via
  vector-boson fusion matched with shower in POWHEG}'',} \textit{ JHEP}
  \textbf{ 02} (2010) 037,
  \href{http://dx.doi.org/10.1007/JHEP02(2010)037}{\doi{10.1007/JHEP02(2010)037}},
\href{http://www.arXiv.org/abs/0911.5299}{\texttt{arXiv:0911.5299}}.

\bibitem{Hartanto:2015uka}
\hrefCMSnoop {}{H.~B. Hartanto, B.~Jager, L.~Reina, and D.~Wackeroth, ``{Higgs
  boson production in association with top quarks in the POWHEG BOX}'',}
  \textit{ Phys. Rev. D} \textbf{ 91} (2015) 094003,
  \href{http://dx.doi.org/10.1103/PhysRevD.91.094003}{\doi{10.1103/PhysRevD.91.094003}},
  \href{http://www.arXiv.org/abs/1501.04498}{\texttt{arXiv:1501.04498}}.

\bibitem{Campbell:2012am}
J.~M. Campbell\hrefCMSnoop {}{ {et~al.}, ``{NLO Higgs boson production plus one
  and two jets using the POWHEG BOX, MadGraph4 and MCFM}'',} \textit{ JHEP}
  \textbf{ 07} (2012) 092,
  \href{http://dx.doi.org/10.1007/JHEP07(2012)092}{\doi{10.1007/JHEP07(2012)092}},
\href{http://www.arXiv.org/abs/1202.5475}{\texttt{arXiv:1202.5475}}.

\bibitem{Luisoni:2013kna}
\hrefCMSnoop {}{G.~Luisoni, P.~Nason, C.~Oleari, and F.~Tramontano,
  ``{$\PH\PW^{\pm}/\PH\PZ$ + 0 and 1 jet at NLO with the POWHEG BOX interfaced
  to GoSam and their merging within MiNLO}'',} \textit{ JHEP} \textbf{ 10}
  (2013) 083,
  \href{http://dx.doi.org/10.1007/JHEP10(2013)083}{\doi{10.1007/JHEP10(2013)083}},
\href{http://www.arXiv.org/abs/1306.2542}{\texttt{arXiv:1306.2542}}.

\bibitem{Gao:2010qx}
Y.~Gao\hrefCMSnoop {}{ {et~al.}, ``{Spin determination of single-produced
  resonances at hadron colliders}'',} \textit{ Phys. Rev. D} \textbf{ 81}
  (2010) 075022,
  \href{http://dx.doi.org/10.1103/PhysRevD.81.075022}{\doi{10.1103/PhysRevD.81.075022}},
  \href{http://www.arXiv.org/abs/1001.3396}{\texttt{arXiv:1001.3396}}.
[Erratum: \DOI{10.1103/PhysRevD.81.079905}].

\bibitem{Frederix:2012ps}
\hrefCMSnoop {}{R.~Frederix and S.~Frixione, ``{Merging meets matching in
  MC@NLO}'',} \textit{ JHEP} \textbf{ 12} (2012) 061,
  \href{http://dx.doi.org/10.1007/JHEP12(2012)061}{\doi{10.1007/JHEP12(2012)061}},
\href{http://www.arXiv.org/abs/1209.6215}{\texttt{arXiv:1209.6215}}.

\bibitem{Grazzini:2015hta}
\hrefCMSnoop {}{M.~Grazzini, S.~Kallweit, and D.~Rathlev, ``{ZZ production at
  the LHC: fiducial cross sections and distributions in NNLO QCD}'',} \textit{
  Phys. Lett.} \textbf{ 750} (2015) 407,
  \href{http://dx.doi.org/10.1016/j.physletb.2015.09.055}{\doi{10.1016/j.physletb.2015.09.055}},
\href{http://www.arXiv.org/abs/1507.06257}{\texttt{arXiv:1507.06257}}.

\bibitem{MCFM}
\hrefCMSnoop {}{J.~M. Campbell and R.~K. Ellis, ``{MCFM for the Tevatron and
  the LHC}'',} \textit{ Nucl. Phys. Proc. Suppl.} \textbf{ 205} (2010) 10,
  \href{http://dx.doi.org/10.1016/j.nuclphysbps.2010.08.011}{\doi{10.1016/j.nuclphysbps.2010.08.011}},
\href{http://www.arXiv.org/abs/1007.3492}{\texttt{arXiv:1007.3492}}.

\bibitem{Campbell:2013una}
\hrefCMSnoop {}{J.~M. Campbell, R.~K. Ellis, and C.~Williams, ``{Bounding the
  Higgs width at the LHC using full analytic results for $gg \to \Pe^- \Pe^+
  \mu^- \mu^+$}'',} \textit{ JHEP} \textbf{ 04} (2014) 060,
  \href{http://dx.doi.org/10.1007/JHEP04(2014)060}{\doi{10.1007/JHEP04(2014)060}},
\href{http://www.arXiv.org/abs/1311.3589}{\texttt{arXiv:1311.3589}}.

\bibitem{Bonvini:1304.3053}
M.~Bonvini\hrefCMSnoop {}{ {et~al.}, ``{Signal-background interference effects
  in $gg \to \PH \to \PW\PW$ beyond leading order}'',} \textit{ Phys. Rev. D}
  \textbf{ 88} (2013) 034032,
  \href{http://dx.doi.org/10.1103/PhysRevD.88.034032}{\doi{10.1103/PhysRevD.88.034032}},
\href{http://www.arXiv.org/abs/1304.3053}{\texttt{arXiv:1304.3053}}.

\bibitem{Catani:2007vq}
\hrefCMSnoop {}{S.~Catani and M.~Grazzini, ``{An NNLO subtraction formalism in
  hadron collisions and its application to Higgs boson production at the
  LHC}'',} \textit{ Phys. Rev. Lett.} \textbf{ 98} (2007) 222002,
  \href{http://dx.doi.org/10.1103/PhysRevLett.98.222002}{\doi{10.1103/PhysRevLett.98.222002}},
\href{http://www.arXiv.org/abs/hep-ph/0703012}{\texttt{arXiv:hep-ph/0703012}}.

\bibitem{Grazzini:2008tf}
\hrefCMSnoop {}{M.~Grazzini, ``{NNLO predictions for the Higgs boson signal in
  the $\PH \rightarrow \PW\PW \rightarrow \ell \nu \ell \nu$ and $\PH
  \rightarrow \PZ\PZ \rightarrow 4\ell$ decay channels}'',} \textit{ JHEP}
  \textbf{ 02} (2008) 043,
  \href{http://dx.doi.org/10.1088/1126-6708/2008/02/043}{\doi{10.1088/1126-6708/2008/02/043}},
\href{http://www.arXiv.org/abs/0801.3232}{\texttt{arXiv:0801.3232}}.

\bibitem{Grazzini:2013mca}
\hrefCMSnoop {}{M.~Grazzini and H.~Sargsyan, ``{Heavy-quark mass effects in
  Higgs boson production at the LHC}'',} \textit{ JHEP} \textbf{ 09} (2013)
  129,
  \href{http://dx.doi.org/10.1007/JHEP09(2013)129}{\doi{10.1007/JHEP09(2013)129}},
\href{http://www.arXiv.org/abs/1306.4581}{\texttt{arXiv:1306.4581}}.

\bibitem{Sjostrand:2015}
T.~Sj{\"o}strand\hrefCMSnoop {}{ {et~al.}, ``An introduction to {PYTHIA}
  8.2'',} \textit{ Comput. Phys. Commun.} \textbf{ 191} (2015) 159,
  \href{http://dx.doi.org/10.1016/j.cpc.2015.01.024}{\doi{10.1016/j.cpc.2015.01.024}},
\href{http://www.arXiv.org/abs/1410.3012}{\texttt{arXiv:1410.3012}}.

\bibitem{Khachatryan:2015pea}
\hrefCMSnoop {}{{CMS Collaboration}, ``Event generator tunes obtained from
  underlying event and multiparton scattering measurements'',} \textit{ Eur.
  Phys. J. C} \textbf{ 76} (2016) 155,
  \href{http://dx.doi.org/10.1140/epjc/s10052-016-3988-x}{\doi{10.1140/epjc/s10052-016-3988-x}},
\href{http://www.arXiv.org/abs/1512.00815}{\texttt{arXiv:1512.00815}}.

\bibitem{CP5}
\hrefCMSnoop {}{{CMS Collaboration}, ``{Extraction and validation of a new set
  of CMS PYTHIA8 tunes from underlying-event measurements}'',} \textit{ Eur.
  Phys. J. C} \textbf{ 80} (2020) 4,
  \href{http://dx.doi.org/10.1140/epjc/s10052-019-7499-4}{\doi{10.1140/epjc/s10052-019-7499-4}},
  \href{http://www.arXiv.org/abs/1903.12179}{\texttt{arXiv:1903.12179}}.

\bibitem{NNPDF2015}
\hrefCMSnoop {}{{NNPDF} Collaboration, ``Parton distributions for the {LHC} run
  {II}'',} \textit{ JHEP} \textbf{ 04} (2015) 040,
  \href{http://dx.doi.org/10.1007/JHEP04(2015)040}{\doi{10.1007/JHEP04(2015)040}},
\href{http://www.arXiv.org/abs/1410.8849}{\texttt{arXiv:1410.8849}}.

\bibitem{GEANT}
\hrefCMSnoop {}{{GEANT4} Collaboration, ``{GEANT4}---a simulation toolkit'',}
  \textit{ Nucl. Instrum. Meth. A} \textbf{ 506} (2003) 250,
\href{http://dx.doi.org/10.1016/S0168-9002(03)01368-8}{\doi{10.1016/S0168-9002(03)01368-8}}.

\bibitem{Geant2}
\hrefCMSnoop {}{J.~Allison {et~al.}, ``{GEANT4 developments and
  applications}'',} \textit{ IEEE Trans. Nucl. Sci.} \textbf{ 53} (2006) 270,
  \href{http://dx.doi.org/10.1109/TNS.2006.869826}{\doi{10.1109/TNS.2006.869826}}.

\bibitem{CMS-PRF-14-001}
\hrefCMSnoop {}{{CMS Collaboration}, ``{Particle-flow reconstruction and global
  event description with the CMS detector}'',} \textit{ JINST} \textbf{ 12}
  (2017) P10003,
  \href{http://dx.doi.org/10.1088/1748-0221/12/10/P10003}{\doi{10.1088/1748-0221/12/10/P10003}},
  \href{http://www.arXiv.org/abs/1706.04965}{\texttt{arXiv:1706.04965}}.

\bibitem{CMS-TDR-15-02}
\href {https://cds.cern.ch/record/2020886}{{CMS Collaboration}, ``Technical
  proposal for the {Phase-II} upgrade of the {Compact Muon Solenoid}'',} CMS
  Technical Proposal CERN-LHCC-2015-010, CMS-TDR-15-02, 2015.

\bibitem{Khachatryan:2015hwa}
\hrefCMSnoop {}{{CMS Collaboration}, ``{Electron and photon reconstruction and
  identification with the CMS experiment at the CERN LHC}'',} \textit{ JINST}
  \textbf{ 16} (2021) P05014,
  \href{http://dx.doi.org/10.1088/1748-0221/16/05/P05014}{\doi{10.1088/1748-0221/16/05/P05014}},
  \href{http://www.arXiv.org/abs/2012.06888}{\texttt{arXiv:2012.06888}}.

\bibitem{Sirunyan:2018fpa}
\hrefCMSnoop {}{{CMS Collaboration}, ``{Performance of the CMS muon detector
  and muon reconstruction with proton-proton collisions at $\sqrt{s}=$ 13
  TeV}'',} \textit{ JINST} \textbf{ 13} (2018) P06015,
  \href{http://dx.doi.org/10.1088/1748-0221/13/06/P06015}{\doi{10.1088/1748-0221/13/06/P06015}},
  \href{http://www.arXiv.org/abs/1804.04528}{\texttt{arXiv:1804.04528}}.

\bibitem{CMS:2020ebo}
\hrefCMSnoop {}{{CMS Collaboration}, ``{Pileup mitigation at CMS in 13 TeV
  data}'',} \textit{ JINST} \textbf{ 15} (2020) P09018,
  \href{http://dx.doi.org/10.1088/1748-0221/15/09/P09018}{\doi{10.1088/1748-0221/15/09/P09018}},
  \href{http://www.arXiv.org/abs/2003.00503}{\texttt{arXiv:2003.00503}}.

\bibitem{tagNprobe}
\hrefCMSnoop {}{{CMS Collaboration}, ``Measurement of the inclusive {$\PW$} and
  {$\PZ$} production cross sections in pp collisions at $\sqrt{s}=7$ {TeV} with
  the {CMS} experiment'',} \textit{ JHEP} \textbf{ 10} (2011) 132,
  \href{http://dx.doi.org/10.1007/JHEP10(2011)132}{\doi{10.1007/JHEP10(2011)132}},
\href{http://www.arXiv.org/abs/1107.4789}{\texttt{arXiv:1107.4789}}.

\bibitem{Cacciari:2008gp}
\hrefCMSnoop {}{M.~Cacciari, G.~P. Salam, and G.~Soyez, ``The
  anti-$k_\mathrm{t}$ jet clustering algorithm'',} \textit{ JHEP} \textbf{ 04}
  (2008) 063,
  \href{http://dx.doi.org/10.1088/1126-6708/2008/04/063}{\doi{10.1088/1126-6708/2008/04/063}},
  \href{http://www.arXiv.org/abs/0802.1189}{\texttt{arXiv:0802.1189}}.

\bibitem{Cacciari:2011ma}
\hrefCMSnoop {}{M.~Cacciari, G.~P. Salam, and G.~Soyez, ``{FastJet user
  manual}'',} \textit{ Eur. Phys. J. C} \textbf{ 72} (2012) 1896,
  \href{http://dx.doi.org/10.1140/epjc/s10052-012-1896-2}{\doi{10.1140/epjc/s10052-012-1896-2}},
\href{http://www.arXiv.org/abs/1111.6097}{\texttt{arXiv:1111.6097}}.

\bibitem{CMS-PAS-JME-16-003}
\href {https://cds.cern.ch/record/2256875}{{CMS Collaboration}, ``{Jet
  algorithms performance in 13\TeV data}'',} CMS Physics Analysis Summary
  CMS-PAS-JME-16-003, 2017.

\bibitem{JINST6}
\hrefCMSnoop {}{{CMS Collaboration}, ``Determination of jet energy calibration
  and transverse momentum resolution in {CMS}'',} \textit{ JINST} \textbf{ 6}
  (2011) P11002,
  \href{http://dx.doi.org/10.1088/1748-0221/6/11/P11002}{\doi{10.1088/1748-0221/6/11/P11002}},
\href{http://www.arXiv.org/abs/1107.4277}{\texttt{arXiv:1107.4277}}.

\bibitem{Khachatryan:2016kdb}
\hrefCMSnoop {}{{CMS Collaboration}, ``{Jet energy scale and resolution in the
  CMS experiment in pp collisions at 8\TeV}'',} \textit{ JINST} \textbf{ 12}
  (2017) P02014,
  \href{http://dx.doi.org/10.1088/1748-0221/12/02/P02014}{\doi{10.1088/1748-0221/12/02/P02014}},
  \href{http://www.arXiv.org/abs/1607.03663}{\texttt{arXiv:1607.03663}}.

\bibitem{JESCalib}
\href {https://cds.cern.ch/record/2715872}{{CMS Collaboration}, ``{Jet energy
  scale and resolution performance with 13 TeV data collected by CMS in
  2016-2018}'',} Detectors and Experimental Techniques CERN-CMS-DP-2020-019,
  2020.

\bibitem{BTAG2018}
\hrefCMSnoop {}{{CMS Collaboration}, ``{Identification of heavy-flavour jets
  with the {CMS} detector in pp collisions at 13 {TeV}}'',} \textit{ JINST}
  \textbf{ 13} (2018) P05011,
  \href{http://dx.doi.org/10.1088/1748-0221/13/05/p05011}{\doi{10.1088/1748-0221/13/05/p05011}},
  \href{http://www.arXiv.org/abs/1712.07158}{\texttt{arXiv:1712.07158}}.

\bibitem{Hocker:2007ht}
\hrefCMSnoop {}{H.~Voss, A.~H{\"o}cker, J.~Stelzer, and F.~Tegenfeldt,
  ``{TMVA}, the toolkit for multivariate data analysis with {ROOT}'',} in
  \textit{ {XIth} International Workshop on Advanced Computing and Analysis
  Techniques in Physics Research ({ACAT})}, p.~40.
\newblock 2007.
\newblock
  \href{http://www.arXiv.org/abs/physics/0703039}{\texttt{arXiv:physics/0703039}}.
\newblock {[PoS(ACAT)040]}.
\href{http://dx.doi.org/10.22323/1.050.0040}{\doi{10.22323/1.050.0040}}.

\bibitem{CombineManual}
\hrefCMSnoop {}{M.~Baak, S.~Gadatsch, R.~Harrington, and W.~Verkerke,
  ``{Interpolation between multi-dimensional histograms using a new non-linear
  moment morphing method}'',} \textit{ Nucl. Instrum. Meth. A} \textbf{ 771}
  (2015) 39,
  \href{http://dx.doi.org/10.1016/j.nima.2014.10.033}{\doi{10.1016/j.nima.2014.10.033}},
  \href{http://www.arXiv.org/abs/1410.7388}{\texttt{arXiv:1410.7388}}.

\bibitem{CMS-LUM-17-003}
\hrefCMSnoop {}{{CMS Collaboration}, ``Precision luminosity measurement in
  proton-proton collisions at ${\sqrt{s}=13\TeV}$ in 2015 and 2016 at {CMS}'',}
  \textit{ Eur. Phys. J. C} \textbf{ 81} (2021) 800,
  \href{http://dx.doi.org/10.1140/epjc/s10052-021-09538-2}{\doi{10.1140/epjc/s10052-021-09538-2}},
  \href{http://www.arXiv.org/abs/2104.01927}{\texttt{arXiv:2104.01927}}.

\bibitem{CMS-PAS-LUM-17-004}
\href {http://cds.cern.ch/record/2621960}{{CMS Collaboration}, ``{CMS}
  luminosity measurement for the 2017 data-taking period at
  $\sqrt{s}={13\TeV}$'',} CMS Physics Analysis Summary CMS-PAS-LUM-17-004,
  2018.

\bibitem{CMS-PAS-LUM-18-002}
\href {http://cds.cern.ch/record/2676164}{{CMS Collaboration}, ``{CMS}
  luminosity measurement for the 2018 data-taking period at
  $\sqrt{s}={13\TeV}$'',} CMS Physics Analysis Summary CMS-PAS-LUM-18-002,
  2019.

\bibitem{Cepeda:2019klc}
\hrefCMSnoop {}{M.~Cepeda {et~al.}, ``{Report from Working Group 2}: {Higgs
  Physics at the HL-LHC and HE-LHC}''}.
\newblock 2019.
\newblock
  \href{http://www.arXiv.org/abs/1902.00134}{\texttt{arXiv:1902.00134}}.
\newblock
  \href{http://dx.doi.org/10.23731/CYRM-2019-007.221}{\doi{10.23731/CYRM-2019-007.221}}.

\bibitem{Read:2002hq}
\hrefCMSnoop {}{A.~L. Read, ``Presentation of search results: the
  {CL}$_\mathrm{s}$ technique'',} in \textit{ Durham IPPP Workshop: Advanced
  Statistical Techniques in Particle Physics}, p.~2693.
\newblock Durham, UK, March, 2002.
\newblock [J. Phys. G 28 (2002) 2693].
\href{http://dx.doi.org/10.1088/0954-3899/28/10/313}{\doi{10.1088/0954-3899/28/10/313}}.

\bibitem{CLS2}
\hrefCMSnoop {}{T.~Junk, ``{Confidence level computation for combining searches
  with small statistics}'',} \textit{ Nucl. Instrum. Meth. A} \textbf{ 434}
  (1999) 435,
  \href{http://dx.doi.org/10.1016/S0168-9002(99)00498-2}{\doi{10.1016/S0168-9002(99)00498-2}},
  \href{http://www.arXiv.org/abs/hep-ex/9902006}{\texttt{arXiv:hep-ex/9902006}}.

\bibitem{CLSA}
\hrefCMSnoop {}{G.~Cowan, K.~Cranmer, E.~Gross, and O.~Vitells, ``Asymptotic
  formulae for likelihood-based tests of new physics'',} \textit{ Eur. Phys. J.
  C} \textbf{ 71} (2011) 1554,
  \href{http://dx.doi.org/10.1140/epjc/s10052-011-1554-0}{\doi{10.1140/epjc/s10052-011-1554-0}},
  \href{http://www.arXiv.org/abs/1007.1727}{\texttt{arXiv:1007.1727}}.
[Erratum: \DOI{10.1140/epjc/s10052-013-2501-z}].

\bibitem{CMS-NOTE-2011-005}
\href {http://cdsweb.cern.ch/record/1379837}{{ATLAS and CMS Collaborations},
  ``Procedure for the {LHC} {H}iggs boson search combination in summer 2011'',}
  Technical Report CMS-NOTE-2011-005, ATL-PHYS-PUB-2011-11, 2011.

\bibitem{Dall'Osso:2015aia}
A.~Carvalho\hrefCMSnoop {}{ {et~al.}, ``{Higgs pair production: choosing
  benchmarks with cluster analysis}'',} \textit{ JHEP} \textbf{ 04} (2016) 126,
  \href{http://dx.doi.org/10.1007/JHEP04(2016)126}{\doi{10.1007/JHEP04(2016)126}},
\href{http://www.arXiv.org/abs/1507.02245}{\texttt{arXiv:1507.02245}}.

\bibitem{hepData}
\hrefCMSnoop {}{{CMS Collaboration}}. {HEPD}ata record for this analysis, 2022.
\newblock
  \href{http://dx.doi.org/10.17182/hepdata.130957}{\doi{10.17182/hepdata.130957}}.

\end{thebibliography}\endgroup

\cleardoublepage \appendix\section{The CMS Collaboration \label{app:collab}}\begin{sloppypar}\hyphenpenalty=5000\widowpenalty=500\clubpenalty=5000
\cmsinstitute{Yerevan Physics Institute, Yerevan, Armenia}
{\tolerance=6000
A.~Tumasyan\cmsAuthorMark{1}\cmsorcid{0009-0000-0684-6742}
\par}
\cmsinstitute{Institut f\"{u}r Hochenergiephysik, Vienna, Austria}
{\tolerance=6000
W.~Adam\cmsorcid{0000-0001-9099-4341}, J.W.~Andrejkovic, T.~Bergauer\cmsorcid{0000-0002-5786-0293}, S.~Chatterjee\cmsorcid{0000-0003-2660-0349}, K.~Damanakis\cmsorcid{0000-0001-5389-2872}, M.~Dragicevic\cmsorcid{0000-0003-1967-6783}, A.~Escalante~Del~Valle\cmsorcid{0000-0002-9702-6359}, P.S.~Hussain\cmsorcid{0000-0002-4825-5278}, M.~Jeitler\cmsAuthorMark{2}\cmsorcid{0000-0002-5141-9560}, N.~Krammer\cmsorcid{0000-0002-0548-0985}, L.~Lechner\cmsorcid{0000-0002-3065-1141}, D.~Liko\cmsorcid{0000-0002-3380-473X}, I.~Mikulec\cmsorcid{0000-0003-0385-2746}, P.~Paulitsch, F.M.~Pitters, J.~Schieck\cmsAuthorMark{2}\cmsorcid{0000-0002-1058-8093}, R.~Sch\"{o}fbeck\cmsorcid{0000-0002-2332-8784}, D.~Schwarz\cmsorcid{0000-0002-3821-7331}, S.~Templ\cmsorcid{0000-0003-3137-5692}, W.~Waltenberger\cmsorcid{0000-0002-6215-7228}, C.-E.~Wulz\cmsAuthorMark{2}\cmsorcid{0000-0001-9226-5812}
% ...
\par}
\vskip\cmsinstskip
\dag:~Deceased\\
$^{1}$Also at Yerevan State University, Yerevan, Armenia\\
$^{2}$Also at TU Wien, Vienna, Austria\\
% ...
\end{sloppypar}
\end{document}